\documentclass[iop]{emulateapj}                             
\newcommand{\figscaleone}{\epsscale{1.15}}      

\slugcomment{Submitted to ApJ 16th January 2013}

\shorttitle{MSP cyclic spectroscopy}
\shortauthors{Walker, Demorest \&\ van~Straten}

\begin{document}

\title{Cyclic spectroscopy of The Millisecond Pulsar, B1937+21}

\author{Mark A. Walker}
\affil{Manly Astrophysics, 3/22 Cliff Street, Manly 2095, Australia}
\email{Mark.Walker@manlyastrophysics.org}
\author{Paul B. Demorest}
\affil{National Radio Astronomy Observatory, Charlottesville, VA 22903, USA}
\email{pdemores@nrao.edu}
\author{Willem van~Straten}
\affil{Swinburne University of Technology, Astrophysics and Supercomputing, Hawthorn 3122, Australia}
\email{willem@swin.edu.au}

\begin{abstract}
Cyclic spectroscopy is a signal processing technique that was originally developed for engineering applications and has recently been introduced into the field of pulsar astronomy. It is a powerful technique with many attractive features, not least of which is the explicit rendering of information about the relative phases in any filtering imposed on the signal, thus making holography a more straightforward proposition. Here we present methods for determining optimum estimates of both the filter itself and the statistics of the unfiltered signal, starting from a measured cyclic spectrum. In the context of radio pulsars these quantities tell us the impulse response of the interstellar medium and the intrinsic pulse profile. We  demonstrate our techniques by application to 428$\,$MHz Arecibo data on the millisecond pulsar B1937+21, obtaining the pulse profile free from the effects of  interstellar scattering. As expected, the  intrinsic profile exhibits main- and inter-pulse components that are narrower than they appear in the scattered profile; it also manifests some weak, but sharp features that are revealed for the first time at low frequency. We determine the structure of the received electric-field envelope as a function of delay and Doppler-shift. Our delay-Doppler image has a high dynamic-range and displays some pronounced, low-level power concentrations at large delays. These concentrations imply strong clumpiness in the ionized interstellar medium, on AU size-scales, which must adversely affect the timing of B1937+21.
\end{abstract}


\section{Introduction}
Although radio pulsars emit intrinsically broad-band radiation, spectroscopy of these sources often reveals a great deal of narrow-band structure (e.g. Rickett 1990). This structure arises during propagation of the signal through the interstellar medium (ISM), where it is scattered by inhomogeneities in the ionized gas --- it is interference between the various scattered waves which creates the observed fringes. Consequently high resolution spectroscopy of pulsars has proved to be a powerful tool for investigating the ISM (Roberts and Ables 1982; Cordes and Wolszczan 1985; Stinebring et al 2001).

Traditionally pulsar spectroscopy is undertaken by forming the power-spectrum of the signal in a pulse-phase window where the flux is high (i.e. ``on-pulse''), and subtracting the power-spectrum from a window where the flux is low (``off-pulse''), so as to remove the steady, background power level. Recently Demorest (2011) has drawn attention to an alternative approach, known as cyclic spectroscopy, in which one measures the modulation of the spectrum across the entire pulse profile. Cyclic spectroscopy was developed in engineering disciplines  for studying signals whose statistics are  periodically modulated (Gardner 1992; Antoni 2007). Signals of this type are common and are referred to as cyclo-stationary. The electric field received from a radio pulsar can be thought of as periodically-amplitude-modulated noise (Rickett 1975),  so radio pulsars provide an example of a signal which is cyclo-stationary.

As described by Demorest (2011), cyclic spectroscopy has several advantages over the simpler method of differencing on-pulse and off-pulse power spectra. Periodic amplitude modulation of the pulsar's radio-frequency noise, introduced by rotation of the pulsar beam, splits the received signal into upper- and lower-sidebands. By construction, the cyclic spectrum is the product of the lower sideband with the complex conjugate of the upper sideband. It is thus a complex quantity and as such it explicitly manifests information about the phase of any filtering which has occured prior to reception. For radio pulsars observed at low frequencies the dominant filtering is due to the ISM --- specifically, to dispersion and scattering of the waves. Thus a time-domain representation of the filter is, to a good approximation, just the impulse-response of the ISM.

In the present paper we show how to determine the filter given a measured cyclic spectrum.  We also show how to determine the intrinsic cyclic spectrum of the signal --- in other words the (Fourier Transform of the) pulse profile which would have been observed in the absence of any scattering or dispersion. These determinations are both made in the narrow-band approximation, appropriate to our data, where there is assumed to be no variation of the intrinsic cyclic spectrum across the observed radio-frequency band. Our main dataset is a 4~MHz bandwidth voltage recording, centered on 428~MHz, of the original  millisecond pulsar, B1937+21 (Backer et al 1982), made with the Arecibo radio telescope.\footnote{The Arecibo Observatory is operated by SRI International under a cooperative agreement with the National Science Foundation (AST-1100968), and in alliance with Ana G. M\'endez-Universidad Metropolitana, and the Universities Space Research Association.}

As far as we are aware the methods presented in this paper are the first attempts to determine both the filter and the intrinsic cyclic spectrum for any astronomical signal. It is possible that our techniques may be useful in fields other than pulsar astronomy, but we do not attempt to identify appropriate fields. Rather we encourage readers to consider applications in other contexts. To aid that process we note here the requirements for validity of our approach: first, the signal must be cyclo-stationary -- i.e. stationary at each phase in its cycle -- in order for the cyclic spectrum to be well-defined. Secondly, our least-squares fitting assumes that the intrinsic cyclic spectrum is just white-noise that is periodically amplitude-modulated, so non-pulsar applications of our techniques are limited to signals which can be described in this or similar fashion. And, finally, the filter must not change significantly within the averaging time over which each cyclic spectrum is constructed. In addition to these requirements, the stopping criterion we employ for our optimizations is based on the assumption of Gaussian noise; but it would be straightforward to modify that criterion. We note that source code is freely available for all the software used herein (see \S5), so readers are free to adapt our code to their purpose.

This paper is organised as follows. In the next section we give some background to the particular problems tackled in this paper. Then in \S3 we show how to determine the filter function and the intrinsic cyclic spectrum by direct construction. In \S4 we consider the issue of optimization --- i.e. obtaining representations of these quantities which best fit the measured cyclic spectrum. In doing so we see that our direct estimate of the intrinsic profile, given in \S3, is in fact the optimum estimate in a least-squares sense. But \S4 does highlight deficiencies in our direct approach to the filter function; so for this quantity we utilize a large-scale optimization of the filter coefficients. Our implementation of this optimization is coded in the language ``C'' and is freely available; it is described in \S5. In \S6 we present results obtained by applying our methods to low-frequency data on PSR B1937+21; both filter functions and intrinsic pulse profiles are presented. Discussion (\S7) and Conclusions (\S8) round out the paper. Two Appendices detail (i) the results of various tests we used to evaluate the code, and (ii) an analysis of  the uncertainties in best-fit parameters. 

\section{Background and general considerations}
Procedures for constructing the cyclic spectrum itself, from a set of recorded voltages, are given by Demorest (2011). We begin our development by quoting the relationship between a signal, $x(t)$, a function of time, with Fourier Transform $X(\nu)$, and the cyclic spectrum of that signal, $S_x$. At modulation frequency $\alpha$ we have (Gardner 1992; Antoni 2007; Demorest 2011)
\begin{equation}
S_x(\alpha,\nu) \equiv \langle X(\nu+\alpha/2)\,  X^*(\nu-\alpha/2)\rangle,
\end{equation}
where the time-average is taken over integer multiples of the period of the system. Thus if we apply a filter, $H(\nu)$, such that the filtered signal is $Z(\nu)=H(\nu)\,X(\nu)$ then the cyclic spectrum of the filtered signal is
\begin{equation}
S_z(\alpha,\nu) = H(\nu+\alpha/2)\,  H^*(\nu-\alpha/2)\, S_x(\alpha,\nu).
\end{equation}
In the case of a radio pulsar the signals $X,Z$ are just electric fields, and the frequency $\nu$ is the radio frequency. Filtering of the signal occurs as a result of propagation, notably dispersion and scattering in the ionized ISM, and in the process of reception, e.g. the bandpass filter. The filter resulting from interstellar propagation evolves on some time-scale, and the average in equation (1) must be restricted to times which are short compared to that evolution time-scale.

Throughout this paper we confine attention to the case of small fractional radio bandwidths, for which we expect the intrinsic cyclic spectrum to be approximately independent of $\nu$:
\begin{equation}
S_x(\alpha,\nu)\rightarrow S_x(\alpha).
\end{equation}
The quantity $S_x(\alpha)$ is already familiar to astronomers from conventional analysis of radio pulsar signals: it is just the Fourier Transform of the pulse profile. But we emphasise that it is the transform of the intrinsic pulse profile, rather than the transform of the measured pulse profile --- the difference being that the latter includes the influence of scattering and other contributions to the filter $H$. 

In general both $S_z$ and $S_x$ are complex quantities, but in the particular case $\alpha=0$ we obtain the zero-modulation-frequency components of the filtered and unfiltered signals, respectively. As these are just the time-averaged  power-spectra of the signals they are non-negative real numbers.

\subsection{Degeneracies}
Before extracting estimates from our data it is necessary to identify and eliminate any degeneracies in the model. Equation (2) shows that there are degeneracies which are multiplicative in form. Writing
\begin{equation}
H(\nu)\rightarrow H(\nu)\,Q(\nu),
\end{equation}
we see that $S_z\rightarrow S_z$ if and only if 
\begin{equation}
S_x(\alpha,\nu)\rightarrow{{S_x(\alpha,\nu)}\over{Q(\nu+\alpha/2)Q^*(\nu-\alpha/2)}}.
\end{equation}
Thus if $S_x$ and $H$ are completely unconstrained then there may be a great deal of degeneracy between these quantities in our model of $S_z$: features seen in the data might be attributed to the intrinsic spectrum or to the effects of an imposed filter. 

In circumstances where the intrinsic cyclic spectrum is independent of radio-frequency (equation 3), the scope of the degeneracy is limited to functions $Q(\nu)$ such that $Q(\nu+\alpha/2)Q^*(\nu-\alpha/2)$ is independent of $\nu$. This condition should hold for all $\alpha$. In the case of small $\alpha$, by expanding to first order in $\alpha$, we see that the form of $Q$ is restricted to those functions satisfying
\begin{equation}
|Q(\nu)|={\rm const.},
\end{equation}
and
\begin{equation}
{{\rm d\;\;}\over{\rm d\nu}}\,{\rm Im}\{\log Q(\nu)\}={\rm const.}
\end{equation}
Hence if we do not know the actual form of $S_x(\alpha)$, then the filter function can only be determined up to an arbitrary multiplicative factor of
\begin{equation}
Q(\nu)=\exp[i(\tau\nu+\phi)+\rho],
\end{equation}
where $\tau$, $\phi$ and $\rho$ are real constants. In other words the overall normalization of $H$, its phase and its phase gradient are all arbitrary, because the simultaneous transformation
\begin{equation}
S_x(\alpha)\rightarrow S_x(\alpha)\exp[-i\tau\alpha-2\rho],
\end{equation}
leaves $S_z$ unchanged. 

If, however, $S_x(\alpha)$ is already known, from previous observations, then the only remaining degeneracy is the overall phase of $H$. This phase is always arbitrary, as can be seen by noting that $\phi$ does not appear in equation (9).

\subsection{Sampling}
For a periodic modulation with period $P=1/\Omega$, as is the case with signals from a radio pulsar, the cyclic spectrum is expected to be zero everywhere except at $\alpha=m\Omega$, where $m$ is an integer, so those are the only modulation frequencies which we sample.  In practice the data are also sampled discretely in the radio-frequency dimension, so we have measurements on a grid, with spacing $\Delta\alpha=\Omega$, and $\Delta\nu$ which we are at liberty to choose. In choosing $\Delta\nu$ the primary consideration relates to structure in the filter function: if we wish to capture signal components which are delayed by times up to $\tau$ then we need to have a resolution $\Delta\nu\le1/2\tau$. One could choose the resolution to be $\ll 1/2\tau$ but that would entail a greater computational load in constructing the cyclic spectrum.

There is a  natural limit to the fineness of the spectral resolution set by $\Delta\nu=\Delta\alpha=\Omega$, corresponding to delays $\tau=\pm P/2$, where $P$ is the pulse period. If there are signal components at delays greater than half the pulse period then the cyclic spectrum is intrinsically undersampled in $\alpha$, because the modulation imposed by the filter function changes significantly on scales $\delta\alpha<\Omega$.

On the other hand there is no difficulty in setting $\Delta\nu \gg \Omega$, providing that there are  no significant signal components at delays greater than $1/\Delta\nu$. 

Although the cyclic spectrum is normally computed on a rectangular grid, values at large $|\alpha|$ and $|\nu|$ may not contain any information. If the voltage signal has a bandwidth $B$, sampled at the Nyquist rate, then the resulting cyclic spectrum is only valid within a diamond-shaped region around the origin, with $|\alpha/2| + |\nu| < B/2$ (Demorest 2011). We also note that there cannot be more information in the cyclic spectrum than was present in the sampled voltage signal from which it was derived. Thus if the cyclic spectrum includes pulse harmonic numbers $m>{\rm N}_p$ (the number of pulses averaged-over), then the pixels in the cyclic spectrum may not be statistically completely  independent. Because of these limitations, the actual number of constraints provided by the data may be smaller than the number of grid points in the cyclic spectrum. 

\subsection{Noise and bias}
The computed cyclic spectrum includes measurement noise which we can characterize in the following way. Suppose that the recorded voltage is $Z(\nu) + N(\nu)$, then we expect the measured cyclic spectrum to be 
\begin{equation}
\langle D(\alpha,\nu)\rangle = S_z(\alpha,\nu) + \langle |N(\nu)|^2\rangle\,\delta(\alpha),
\end{equation}
where the delta-function appears because the measurement noise is stationary. Thus our measured cyclic spectrum is free of noise bias except at $\alpha=0$.

Because modulation is the fundamental characteriztic of pulsar radiation which allows it to be distinguished from measurement noise, estimating the unmodulated part of the cyclic spectrum, $S_z(0,\nu)$, from $D(0,\nu)$ is ambiguous. In this paper we therefore make no attempt to quantify $S_z(0,\nu)$, nor do we make direct use of $D(0,\nu)$ in our estimates of the signal properties $S_x(\alpha)$ and $H(\nu)$.  In turn this means that we are giving up any possibility of determining $S_x(0)$, the zero-frequency term in the Fourier representation of the intrinsic pulse profile. We therefore adopt the convention $S_x(0)=0$ in our models throughout the rest of this paper. 

The actual data which we record, $D(\alpha, \nu)$, will differ from $\langle D\rangle$ because of measurement noise and because the signal itself is stochastic in nature. If there is no averaging (see discussion following equation 14) the variance of the measured cyclic spectrum is given by (Antoni 2007)
\begin{equation}
{\rm var}\{D(\alpha,\nu)\}=\langle D(0,\nu-\alpha/2)\rangle\,\langle D(0,\nu+\alpha/2)\rangle.
\end{equation}
At zero modulation frequency, we recover from equation 11 the familiar result for stationary signals that the variance of the unaveraged power is just the square of the mean power.

For observations of radio pulsars with current instrumentation, noise power is usually the dominant contribution to $D(0,\nu)$ and in this case we have
\begin{equation}
{\rm var}\{D(\alpha,\nu)\}\simeq \langle |N(\nu-\alpha/2)|^2\rangle\, \langle |N(\nu+\alpha/2)|^2\rangle.
\end{equation}
If the measurement noise is white, as is often the case in practice, then equation 12 yields a uniform variance,
\begin{equation}
{\rm var}\{D\}= \langle |N(\nu)|^2\rangle^2 =  \sigma^2, 
\end{equation}
over the entire cyclic spectrum. It is straightforward to estimate $\sigma$, because at zero modulation frequency the cyclic spectrum is just a power spectrum. Thus the noise level is just
\begin{equation}
\sigma ={{ {\cal S}_{sys}}\over\sqrt{\Delta t\, \Delta\nu}},
\end{equation}
where ${\cal S}_{sys}$ is the system-equivalent-flux-density, $\Delta t$ is the integration time and $\Delta\nu$ the channel width. (Here we consider only a single polarization state, but clearly the results can be generalised to different combinations of polarization states.)

Equation 14 clarifies what is meant by the ``no averaging'' requirement immediately preceding equation 11. For cyclic spectroscopy of pulsars the natural choice of spectral resolution is $\Delta\nu=\Delta\alpha$, and we always have $\Delta\alpha=1/P$, where $P$ is the pulse period. Thus for $\Delta t=P$ we have a time-bandwidth product of unity -- a single sample of the signal -- and $\sigma={\cal S}_{sys}$. Equation 11 is then appropriate to a single pulse, and if the cyclic spectrum is averaged over ${\rm N}_p$ pulses the variance is smaller  by a factor $1/{\rm N}_p$.

\section{Direct construction of filter and profile}
We now turn to the task of estimating the filter function (ISM impulse response) and the intrinsic (unscattered) pulse profile starting from a measured cyclic spectrum. We can approach both of these tasks by iteration, as we now describe.
\subsection{Determining the filter function}
Suppose we have a model for $S_x$, but we have incomplete knowledge of $H$. If we know the value of $H$ at a single frequency, $\nu_1$, we can determine its value at nearby frequencies using the measured cyclic spectrum in the vicinity of $\nu_1$, thus:
\begin{eqnarray}
H(\nu_1+\alpha) \simeq{{D(\alpha,\nu_1+\alpha/2)}\over{ H^*(\nu_1)\, S_x(\alpha)}}.\qquad\;\,
\end{eqnarray}
We can make a better estimate of $H$ at a given frequency if we know several nearby values of $H$. Making the replacement $\nu\rightarrow\nu-\alpha/2$ in eq. (2), multiplying by $H(\nu-\alpha)\, S_x^*(\alpha)$ and summing yields
\begin{equation}
H(\nu)={{\sum\limits_{\alpha\ne0} \; D(\alpha,\nu-\alpha/2)\,H(\nu-\alpha)\,S_x^*(\alpha)} \over {\sum\limits_{\alpha\ne0} \; |H(\nu-\alpha)|^2\, |S_x(\alpha)|^2 }},
\end{equation}
where we have used the data, $D$, as our estimate of $S_z$. This equation allows us to construct $H$, in regions where it is unknown, from nearby regions where it has already been determined, providing only that we have formed an estimate of $S_x$. We note that equation (16) includes equation (15) as a special case where $H$ is known only at a single frequency.

Although the development in this section has focused on the idea of obtaining an estimate of $H$ at frequencies where it is not known, it is clear that one could employ equation (16) even if we already have an estimate of $H(\nu)$ for all frequencies, so it can also be viewed as a procedure for updating an existing model of $H$. We will return to this idea in \S3.3 and \S4.

\subsection{Determining the intrinsic spectrum}
Now suppose that we have a model for $H$, what then do the data tell us about $S_x$? Multiplying equation (2) by $H^*(\nu+\alpha/2)\,  H(\nu-\alpha/2)$ and summing over $\nu$ gives
\begin{equation}
S_x(\alpha)={{\sum\limits_\nu \; D(\alpha,\nu)\,H(\nu-\alpha/2)H^*(\nu+\alpha/2)} \over {\sum\limits_\nu \; |H(\nu-\alpha/2)|^2\, |H(\nu+\alpha/2)|^2 }},
\end{equation}
where, again, we have used the data, $D$, as our estimate for $S_z$. Thus, given data and a model for the filter function, we can obtain an estimate of the intrinsic pulse profile implied by the observed cyclic spectrum. Notice that this formula implies a unique estimate of $S_x$ associated with any given pair $D,H$. We shall see in \S4 that equation (17) is actually the optimum estimate of $S_x$, in a least-squares sense, given the data $D$ and the filter $H$.

\subsection{Bootstrap}
From the foregoing we can see that it is straightforward to form an estimate of $H$ given $S_x$, and vice versa. But initially we might not know either. In this situation it is natural to proceeed iteratively, starting with crude estimates and then using equations (16) and (17) repeatedly to improve those estimates. One way of starting the process is to initialize the intrinsic cyclic spectrum to $S_x(\alpha)\leftarrow\langle  D(\alpha,\nu)\rangle_\nu$, i.e. the observed (scattered) pulse profile. This corresponds to the model $H(\nu)=1$ and we could commence iteration of equations (16) and (17) using this approximation. 

Alternatively, having specified our initial estimate of $S_x$ we can build up our estimate of $H$ gradually, using equation (16), starting from an estimate of its value at a single frequency, $H(\nu_1)$. Because the overall phase of $H$ is arbitrary (\S2.1) we are free to choose the phase of $H(\nu_1)$, e.g. phase zero, so only $|H(\nu_1)|$ need be specified in order to start the iteration. One possible initialization is thus $H(\nu_1)\leftarrow\sqrt{|D(\Omega,\nu_1)/S_x(\Omega)|}$, and from there we can gradually build $H$ over the full range of radio frequencies, with information flowing outward from $\nu_1$ towards the edges of the band. In this approach one simply initializes $H$ to zero for frequencies where no estimate has previously been made, so that those frequencies make no contribution to the estimator in equation (16). 

Once this is done we can improve our estimate of the intrinsic cyclic spectrum, $S_x$, by application of equation (17), then we can get a better estimate of $H$ by applying equation (16), and these iterations can be repeated. Thus, if we know neither $S_x$ nor $H$, we can build bootstrap estimates for both of these quantities, given a measured cyclic spectrum. 

The procedure just described is the method which we initially used to solve for $H$ and $S_x$, from the first measured cyclic spectra of a radio pulsar (i.e. the data used in \S5). Broadly speaking the method works: we found that it provided a good representation of much of the structure in the cyclic spectra, and the intrinsic profile was significantly narrower than the scattered profile (see figure 3 in Demorest, 2011). But it did also exhibit some deficiencies, as we describe below. 

\subsection{Deficiencies of the direct method}
One problem which we anticipated is the difficulty of constructing $H$ in regions where $|H|$ is small. In these regions the solution for $H$ is sensitive to noise in the data. In particular it is susceptible to phase jumps at points where $|H|\rightarrow0$: the solutions on either side of the zero can be mutually inconsistent. There are two reasons why this problem arises. One is fundamental: a zero in $|H|$ is an absence of phase information at that particular point, and this cannot be overcome by using different methods of solution. The other reason is specific to the solution method we have presented: the summation in equation (16) includes information coming from both sides of the zero, so each side tries to rotate the phases of the other in order to bring about consistency, but neither side succeeds. In other words, phase discontinuities at zeros of $|H|$ constitute traps for this method of solution. It is not necessary for $|H|$ to be precisely zero in order for a trap to form; it suffices for the signal-to-noise ratio to be low ($\la1$ on a per-channel basis). Trapping was indeed observed in the results we obtained using the approach described above, with significant residuals commonly occuring in the vicinity of points where $|H|$ is small.

It is clearly possible to modify the solution method so as to be less susceptible to these traps. Most obviously, one can restrict the summations in equation (16) to values of $\alpha$ with a single sign --- so that we are only using the information from frequencies $>\nu$ (or $<\nu$) in our estimate for $H(\nu)$. In this scheme information flows in only one direction across the zeros, so one side dictates phase to the other. In practice we observed that this modification did decrease the prevalence of trapping. However, in using only one sign of $\alpha$ we are ignoring half of the information available to constrain $H$ at any given value of $\nu$, so the resulting solution for $H$ cannot be optimum. In the next section we present methods for obtaining the best fit solutions for $H$ and $S_x$.

\section{Optimum estimates of filter and profile}
In estimating $H$ and $S_x$ what we really want are the models which best fit the data, so we have an optimization problem. We introduce the residual between model and data:
\begin{equation}
R(\alpha,\nu)\equiv S_z(\alpha,\nu)-D(\alpha,\nu),
\end{equation}
and we seek to minimize the magnitude of these residuals.

Suppose our data, $D$, have ${\rm N_\nu}$ radio-frequency channels, and ${\rm N_\alpha}$ modulation-frequency bins. In this case we are modelling a filter with ${\rm N_\nu}$ complex unknowns, and an intrinsic cyclic spectrum with ${\rm N_\alpha}/2$ complex unknowns. (The pulse profile is a real quantity, so the spectrum at negative modulation frequencies is simply the complex conjugate of that at positive frequencies.) Thus there are ${\rm N_\nu + N_\alpha/2}$ complex unknowns and $\sim{\rm N_\nu\times N_\alpha/2}$ complex constraints provided by the data, so for ${\rm N_\nu, N_\alpha \gg 1}$ the model is over-determined. In this situation we cannot make the residuals zero everywhere and we simply aim to make them small.

Here we follow the usual practice of minimizing the sum-of-squares of the residuals
\begin{equation}
M\equiv\sum\limits_{\nu,\alpha\ne0}R^*\,R,
\end{equation}
with respect to all of the model parameters. We then have
\begin{equation}
{{\partial M}\over{\partial q}} =2\sum\limits_{\nu,\alpha\ne0} R\,{{\partial S_z^*}\over{\partial q}},
\end{equation}
where $q$ represents any of the model parameters which define $H$ and $S_x$, and minimization of $M$ implies
\begin{equation}
{{\partial M}\over{\partial q}} =0
\end{equation}
for every parameter $q$. 

We compute the derivative for each parameter in turn. Each value of $H$ and $S_x$ is complex and thus involves two distinct real parameters. We take these to be the real and imaginary parts of the coefficients.  For $S_{rm}:={\rm Re}\{S_x(\alpha_m)\}$, $S_{im}:={\rm Im}\{S_x(\alpha_m)\}$, we have
\begin{equation}
{{\partial M}\over{\partial S_{rm}}} + i {{\partial M}\over{\partial S_{im}}} = \nabla_{\!S} M\big|_{\alpha=\alpha_m},
\end{equation}
where
\begin{equation}
 \nabla_{\!S} M :=4\sum\limits_{\nu}R(\alpha,\nu)\,H(\nu-\alpha/2)\,H^*(\nu+\alpha/2).
\end{equation}
And for $H_{rk}:={\rm Re}\{H(\nu_k)\}$, $H_{ik}:={\rm Im}\{H(\nu_k)\}$, we have
\begin{equation}
{{\partial M}\over{\partial H_{rk}}} + i {{\partial M}\over{\partial H_{ik}}} = \nabla_{\!H} M\big|_{\nu=\nu_k},
\end{equation}
where
\begin{equation}
\nabla_{\!H} M :=4\sum\limits_{\alpha\ne0}R(\alpha,\nu-\alpha/2)\, H(\nu-\alpha)\,S_x^*(\alpha).
\end{equation}

Having determined a demerit function, $M$, and the gradient of $M$ with respect to each of the parameters of interest, we are in a position to employ one of various standard methods (e.g. Nocedal and Wright 1999)  to the problem of optimizing our solutions. Before turning to the choice of method, and the details of its application, it is helpful to establish the relationship between our ``direct'' solutions of \S3 and the optimum estimates which we are seeking.

\subsection{Relationship of direct solution to least-squares}
We have already noted (\S3) that our ``direct'' procedure for constructing $H$ -- i.e. equation (16) -- could also be regarded as an algorithm for updating $H$, given an existing estimate. Explicitly, the update is $H\rightarrow H+\Delta H$, where
\begin{equation}
\Delta H(\nu):=-{{\sum\limits_{\alpha\ne0} \; R(\alpha,\nu-\alpha/2)\,H(\nu-\alpha)\, S_x^*(\alpha)} \over {\sum\limits_{\alpha\ne0} \; |H(\nu-\alpha)|^2\, |S_x(\alpha)|^2 }}.
\end{equation}
We can also rewrite equation (17) as an update for the intrinsic spectrum, $S_x\rightarrow S_x+\Delta S_x$, with
\begin{equation}
\Delta S_x(\alpha):=-{{\sum\limits_\nu \; R(\alpha,\nu)\,H(\nu-\alpha/2)H^*(\nu+\alpha/2)} \over {\sum\limits_\nu \; |H(\nu-\alpha/2)|^2\, |H(\nu+\alpha/2)|^2 }}.
\end{equation}
In both cases we recognise the numerator to be (up to a constant factor) just the gradient of $-M$ with respect to the corresponding parameters. This is comforting because it suggests that our ``direct'' method is moving the estimates in a direction which will improve the model. To be confident that this is the case we need to gauge the step-size, not just its direction, and to achieve that it is helpful to evaluate the second derivatives of $M$.

The curvature of $M$ with respect to our various parameters is given by differentiating equations 23 and 25. The results are
\begin{eqnarray}
{{\partial^2 M}\over{\partial S_{rm}^2}} =4\,\sum\limits_\nu \; |H(\nu+\alpha_m/2)|^2\, |H(\nu-\alpha_m/2)|^2,\cr
={{\partial^2 M}\over{\partial S_{im}^2}},\qquad\qquad\qquad\qquad\qquad\qquad\qquad
\end{eqnarray}
and
\begin{eqnarray}
{{\partial^2 M}\over{\partial H_{rk}^2}} =4\,\sum\limits_{\alpha\ne0} \; |H(\nu_k-\alpha)|^2\, |S_x(\alpha)|^2,\cr
={{\partial^2 M}\over{\partial H_{ik}^2}}.\qquad\qquad\qquad\qquad\quad\;\;
\end{eqnarray}
We can now see that for each of our real parameters, $q$, the ``direct'' estimate in \S3 is an iteration with updates (equations 26, 27) $\Delta q$:
\begin{equation}
\Delta q =  -\left[ {{\partial^2 M}\over{\partial q^2}}\right]^{-1} {{\partial M}\over{\partial q}}.
\end{equation}
This form is just Newton's method applied to each parameter separately. Equivalently: it is a simultaneous multi-parameter quasi-Newton method in which the off diagonal elements of the Hessian are neglected.

We can check whether or not this is a good approximation to the actual Hessian by explicitly computing the off-diagonal terms. In the case where both $q_i$ and $q_j$ relate to $S_x$ these off-diagonal elements are all zero. Furthermore, because the diagonal terms (equation 28) are independent of $S_x$, all of the higher derivatives of $M$ with respect to $S_x$ are zero --- the hypersurface of $M$ is quadratic in $S_x$ when $H$ is fixed. This is no surprise because the residual (equation  18) is linear in $S_x$, and $M$ is quadratic in the residual. It follows that Newton's method yields an exact solution for $S_x$ in a single step. Thus we see that our direct estimate of $S_x$, given in equation (17),  is also the least-squares solution appropriate to the filter $H$ and the data $D$; no additional optimization steps are necessary.

Unfortunately this is not true of the filter function, $H$: neither the off-diagonal elements of the Hessian nor the higher order derivatives are zero in this case. The fact that the off-diagonal terms of the Hessian are non-zero means that we should not expect the filter update (equations 16, 26) to yield a good model. We now turn to the problem of optimizing our model filter function.

\subsection{Optimisation of the filter}
To optimize our model filter we can employ one of the established quasi-Newton methods, in which an approximate (inverse-) Hessian is constructed at each iteration, based on the local properties of the hypersurface $M$ revealed in previous iterations (see, e.g., Nocedal and Wright 1999). A popular choice is the Broyden-Fletcher-Goldfarb-Shanno (BFGS) update, and that is the method we employ in \S5 and subsequently. 

By utilizing a BFGS update to our estimate of $H(\nu)$ we expect to do significantly better than the update in equation 16. This is a general expectation, but it also applies to the particular problems noted in \S3.4: by including the non-zero off-diagonal curvatures of $M$ we provide some of the information needed for the algorithm to escape the traps introduced by zeros in $|H(\nu)|$. However, we ought to be able to do better still if we do not actually seek to construct $H$ in frequency-space, where the traps are localised, but in the Fourier-space conjugate to frequency, i.e. lag-space.

We introduce the lag-space description of the filter, $h(\tau)$, which is related to the frequency-space description of the filter, $H(\nu)$, by the usual Fourier relationships for discretely sampled functions:
\begin{equation}
h_j\equiv h(\tau_j)=\sum\limits_k H_k\exp[2\pi i \tau_j \nu_k],
\end{equation}
and
\begin{equation}
H_k\equiv H(\nu_k)={1\over{N_\nu}}\sum\limits_j h_j\exp[-2\pi i \tau_j \nu_k].
\end{equation}
To optimize our model filter in lag-space we need to know the gradient of $M$ with respect to the lag-space filter coefficients, $h_{rj}:={\rm Re}\{h(\tau_j)\}$ and $h_{ij}:={\rm Im}\{h(\tau_j)\}$. Noting that $\{h_j\}$ and $\{H_k\}$ are different representations of the same information we can write
\begin{equation}
{{\partial M}\over{\partial h_{rj}}} = \sum_k \left\{ {{\partial H_{rk}}\over{\partial h_{rj}}}  {{\partial M}\over{\partial H_{rk}}} + {{\partial H_{ik}}\over{\partial h_{rj}}}  {{\partial M}\over{\partial H_{ik}}}\right\},
\end{equation}
and similarly for the derivative with respect to $h_{ij}$. In this way we find
\begin{equation}
{{\partial M}\over{\partial h_{rj}}} + i {{\partial M}\over{\partial h_{ij}}} = \nabla_{\!h} M\big|_{\tau=\tau_j},
\end{equation}
where
\begin{equation}
\nabla_{\!h} M:={1\over{N_\nu}} \sum\limits_\nu\nabla_{\!H} M\, \exp[2\pi i \tau \nu].
\end{equation}

Similarly one can show that
\begin{equation}
\nabla_{\!H} M = \sum\limits_\tau\nabla_{\!h} M\, \exp[-2\pi i \tau \nu].
\end{equation}
So as an alternative to computing the frequency-space derivatives and determining the lag-space derivatives from them, we can compute the lag-space derivatives first and then determine the frequency-space derivatives. Formally the two different paths to either frequency-space or lag-space derivatives are equivalent. In practice we computed the lag-space derivatives as our primary quantities, using
\begin{equation}
\nabla_{\!h} M\! ={4\over{N_\nu}}\!\!\!\sum\limits_{\nu,\alpha\ne0}\!\!\!R(\alpha,\nu)\,S_x^*(\alpha)H(\nu-\alpha/2)\exp[\pi i \tau_j(2\nu+\alpha)],
\end{equation}
and determined the frequency-space derivatives, if required, using  equation 36. There appears to be no significant difference in computation time between the two approaches.

\subsection{Uncertainties in best-fit parameters}
Suppose that we have obtained our best-fit model, the question then arises ``how accurate is that model?'' To address this issue we need a description of the behaviour of the demerit, $M$, in the vicinity of the best fit. 

At the best-fit point in parameter space, which we denote by $\{q_{jo}\}$, $\nabla_{\!S}M=0$, and either $\nabla_{\!H}M=0$ or $\nabla_{\!h}M=0$. If $M=M_o$ at the best-fit point, then in the immediate neighbourhood of this point the variation of $M$ can be approximated by
\begin{equation}
M\simeq M_o + \sum\limits_{j,m} {1\over2} {{\partial^2 M}\over{\partial q_m\partial q_j}} (q_m-q_{mo}) (q_j-q_{jo}).
\end{equation}
For Gaussian noise the normalized demerit, $M/\sigma^2$, is distributed like $\chi^2$ with ${\rm N_{dof}\simeq (N_\nu - 1)(N_\alpha -2)}$ degrees of freedom, and we expect $M_o\simeq {\rm N_{dof}\sigma^2}$. The fit becomes significantly worse if we move away from the optimum point to any other point such that $M-M_o=\sigma^2$ (Avni 1976), and this contour of $M$ delineates the range of uncertainties in our fit.

Uncertainties in the individual fit parameters can be readily determined if the Hessian, ${{\partial^2 M}/{\partial q_m\partial q_j}}$, is diagonal so that the parameters are all independent of each other. In this case the standard deviation, $\delta q_j$, is given by
\begin{equation}
(\delta q_j)^2= 2\sigma^2 \left({{\partial^2 M}\over{\partial q_j^2}}\right)^{-1}.
\end{equation}
If the Hessian is not diagonal then the parameters are covariant and it is a much more difficult task to describe the uncertainties in the fit. Because we know how $M$ depends on each of the various parameters, we can evaluate the elements of the Hessian explicitly. Doing so we find that the Hessian is indeed diagonal with respect to the set of parameters describing $S_x$, so equation 39 correctly describes the constraints which our model places on those parameters. However, the Hessian is not diagonal with respect to either $\{H_k\}$ or $\{h_j\}$. The standard errors as given by equation 39 are evaluated in an Appendix, while in the next section we discuss parameter covariance.

\subsubsection{Covariances of $\{h_j\}$}
Unfortunately the curvatures given in equation A8 are not the whole story when it comes to describing the uncertainties in the impulse-response function, because there are non-zero off-diagonal elements of the Hessian in respect of these parameters. It is beyond the scope of this paper to give a detailed description of the effect of these mixed curvature terms; here we only draw attention to their significance, deferring a thorough treatment to a later paper.

To illustrate the importance of the off-diagonal elements of the Hessian we employ the simplest filter model, $H(\nu)=1$. In this case we find by direct calculation that in addition to the leading diagonal (described by equation A8), there is a single reverse-diagonal on which the curvatures are non-zero. This reverse-diagonal cuts the leading diagonal at $\tau_m=\tau_j=0$, and for $|\tau_m-\tau_j|\ll w$, the pulse-width, the mixed curvatures are comparable in size to the diagonal elements. The upshot of this is that the combination of complex coefficients $h(\tau_j)+h^*(-\tau_j)$ is tightly constrained, whereas the combination  $h(\tau_j)-h^*(-\tau_j)$ is poorly constrained. The former combination can be thought of as a pure-amplitude modification of the filter $H(\nu)$, whereas the latter is a pure-phase modification. And the fact that these particular combinations of parameters are well-constrained/poorly-constrained for $|\tau_m-\tau_j|\ll w$ is directly attributable to the (in)sensitivity of $H(\nu+\alpha/2)H^*(\nu-\alpha/2)$ to these types of modification.

\section{Implementation of filter optimization}
Having already established that the simple quasi-Newton method of \S3 works tolerably well for our optimization problem, even though all the off-diagonal elements of the Hessian are neglected, our next step is to implement a more sophisticated quasi-Newton method, the BFGS algorithm, to optimize our filter coefficients. More precisely, because of the large number of parameters ($\sim10^4$) needed to describe the filter coefficients, we utilize a ``limited memory'' algorithm, which we call L-BFGS,  in which the full inverse-Hessian is not constructed  (Nocedal 1980; Liu and Nocedal 1989).

We employed the L-BFGS algorithm coded in the {\it NLopt\/} library\footnote{http://ab-initio.mit.edu/nlopt} (Steven G. Johnson, {\it ``The NLopt nonlinear-optimization package''}). The {\it NLopt\/} package was chosen because it is free, portable and offers a wide variety of optimization algorithms (see \S5.3.1). In addition we utilized the {\it FFTW\/} Fourier Transform package\footnote{http://www.fftw.org} from the same group (Frigo and Johnson 2005). Our code is written in C and is freely available.\footnote{https://github.com/demorest/Cyclic-Modelling} It makes use of the PSRCHIVE library\footnote{http://psrchive.sourceforge.net} (Hotan et al 2004; van Straten et al 2012) for file input and therefore can accept data in a variety of formats, including the standard PSRFITS pulsar data format (Hotan et al 2004).

Perhaps the first point to make, here, is that we have chosen to optimize the filter coefficients separately from the parameters which describe our model of the intrinsic cyclic spectrum, $S_x(\alpha)$. There are several reasons for this choice. The strongest motivation is that it allows us to enforce a common timing reference on all our filter solutions, by using the same intrinsic cyclic spectrum throughout. A common timing reference is of paramount importance for all astrophysical studies which rely on pulse arrival-time measurements. Furthermore, by using a common timing (pulse-phase) reference, we can obtain a high signal-to-noise ratio measurement of the intrinsic spectrum by averaging over all our data.

The degeneracies discussed in \S2.1 provide further, minor motivations for separate optimization of filter and intrinsic cyclic spectrum models, as these degeneracies must be eliminated in order for any algorithm to identify the best fit solution. For the overall normalization and phase of the filter that is fairly straightforward, but controlling the degeneracy in phase-gradient is not so easy if both $H$ and $S_x$ are simultaneously adjusted. By contrast, there is no degeneracy in phase-gradient if $S_x$ is fixed.

We noted in \S2.1 that the overall phase of $H$ is always arbitrary, and this degeneracy must be eliminated before we can determine the model filter which best fits the data. We remove this freedom by forcing the imaginary part of  $h(\tau)$ (or $H(\nu)$) to be zero at the point where $|h(\tau)|$ (or $|H(\nu)|$) attains its largest value. Because this choice is arbitrary, once an optimized filter is obtained we are free to rotate its overall phase to any preferred value. If we have a temporal sequence of filters (see \S7.5), the appropriate choice of phase for a given filter is the one which yields the closest match between the current and the previous (or subsequent) filter, leaving only a single, arbitrary phase for the whole temporal sequence.

\subsection{Initialisation}
We make use of two different initializations, which we refer to as ``Unit'' and ``Proximate''. In the case of Unit initialization we begin with $|H(\nu)|=1$, for all radio frequencies, and a constant phase gradient in $H(\nu)$, chosen to match the mean phases seen in the data at $\alpha=\Omega$. For lag-space optimization this initialization corresponds to a delta-function model for $h(\tau)$. Naturally, Unit initialization is only sensible if the overall normalization of our model $S_x(\alpha)$ is consistent with that of the data, $D(\alpha,\nu)$, and we therefore also normalize $S_x(\alpha)$ appropriately.

Unit initialization is appropriate if we have no prior information on the actual structure which is present in the filter function at the time the cyclic spectrum was recorded. Usually there are many cyclic spectra recorded during a single epoch of observation --- e.g. in \S6 we present data from three separate epochs of observation, totalling several hundred cyclic spectra. In such cases the averaging time  for each spectrum is chosen to be small enough that the changes in the filter function between adjacent cyclic spectra are small. Consequently, if we have already optimized the filter appropriate to one cyclic spectrum then that model provides us with a good starting point for modelling the next filter function: that scheme is what we refer to as Proximate initialization. 

\subsection{Stopping criterion}
At what point should we stop the optimization? The {\it NLopt\/} algorithms include various criteria which may be used to recognise that the optimization is complete. Our aim is to find the minimum of $M$, but we do not know ahead of time the precise value of that minimum, so a natural choice of stopping criterion is that $M$ should change by less than a certain, small fractional value during a single iteration of the algorithm. We can determine what that fractional tolerance should be as follows.

In \S2.3 we gave expressions for the variance of $D(\alpha,\nu)$. In particular we noted that ${\rm var}\{D\}=\sigma^2$, a constant, is usually a good approximation in practice. Furthermore, at large modulation frequencies, $\alpha\gg\Omega$, the noise is usually much larger than the signal we're interested in, so it is straightforward to get an estimate of $\sigma^2$ directly from the data. 

For Gaussian noise, which is appropriate to the thermal noise component, we expect the best-fit value of $M$ to conform to a $\chi^2$ distribution, with ${\rm N_{dof}\simeq (N_\nu-1)( N_\alpha  - 2)}$ degrees of freedom. In this case the minimum demerit is expected to be $M_{min}\simeq{\rm N_{dof}\sigma^2}$, and $\sigma^2$ is a significant change in $M$ (Avni 1976), so it is appropriate to stop the optimization once the changes in $M$ are small compared to $\sigma^2$. This translates directly into the requirement that fractional changes in $M$ should be small compared to ${\rm1/N_{dof}}$.  Therefore in this paper the usual stopping criterion is that the fractional change in $M$ should be less than ${\rm0.1/N_{dof}}$.

If the noise is not uniform -- e.g. at the edges of the band, where the instrumental response rolls off, or because of strong Radio Frequency Interference -- one can determine the variance at each point in the cyclic spectrum using equation 11. In this case the residuals (equation 19) should be normalized by the variance at each point $(\alpha,\nu)$ prior to summation. The resulting figure of merit will then be distributed like $\chi^2$. It is straightforward to measure the noise variation across the band, as per \S6.1 (see the top panel in figure 1).

\subsection{Choice of optimization approach}
The various tests described in the Appendix demonstrate that, of the various optimization approaches we tried,  the best method for this problem is L-BFGS in lag-space from Proximate initial conditions; we therefore utilize that method.

\section{Observations of PSR B1937+21}
All of the data utilized in this paper are Arecibo observations of PSR B1937+21 (Backer et al 1982), at radio frequencies close to 428~MHz. The ATNF Pulsar Catalogue\footnote{www.atnf.csiro.au/people/pulsar/psrcat}\ (Manchester et al 2005) reports the following characteriztics for this pulsar: a period of 1.558~ms, a dispersion measure of $71\,{\rm pc\,cm^{-3}}$ (Cognard et al 1995), and a mean 400~MHz flux of 240~mJy (Foster, Fairhead and Backer 1991). Most of the data we use come from a single 4~MHz band centered on 428~MHz, with the exception being an additional 4~MHz chunk, centered on 432~MHz, that we use exclusively in \S6.3 (intrinsic pulse profile determination).  We observed at three different epochs: MJD53791, MJD53847 and MJD53873. Dual-polarization voltages were recorded for intervals of order an hour at each epoch, using the Arecibo Signal Processor baseband recorder (ASP; Demorest 2007), with digitisation at 8-bits per sample. This high dynamic range sampling proved valuable in mitigating the effects of Radio Frequency Interference (\S6.4). We did not attempt any polarization calibration for our data; all the results reported here are based on summing the two polarizations (i.e. the orthogonal feeds of the telescope), as an approximation to Stokes-I.

Individual cyclic spectra were generated from the recorded voltages, using the method described by Demorest (2011). In our first processing of the data we constructed cyclic spectra, averaged over 15 seconds, with 6230 radio-frequency channels and 511 pulse-phase bins. These values were chosen so as to make $\Delta\nu$ as nearly equal to $\Delta\alpha$ as possible, because our first attempts at modelling $H$ and $S_x$ (using the method described in \S3), avoided interpolations. However, the improved fitting method described in \S\S4,5 employs precise interpolation, so it is no longer necessary to match the resolutions in this way. Nor is it preferred, as array sizes which are integer powers of two are better matched to the Fast Fourier Transform algorithm, which we utilize. All the tests of our optimization software, reported in an Appendix, were conducted on the cyclic spectra obtained in our first processing of the data.

Analysis of the cyclic spectra from our first processing revealed some leakage at the edges of the bandpass filter. This is undesirable, particularly because any out-of-band signal is aliased by $\pm4\,$MHz, and will thus appear delayed by approximately $\pm30\,$ms due to incorrect dedispersion. In turn this leaked signal may introduce low-level contamination into our profile estimates or our filter models, or both. We therefore decided to completely reprocess our data, to deal with the leakage and to correct some other minor defects which we were aware of.

In the second processing we produced cyclic spectra averaged over 15 seconds, with 4608 channels and 1024 pulse-phase bins. This reprocessing utilized the cyclic spectrum implementation now freely available as part of the DSPSR software package\footnote{http://dspsr.sourceforge.net} (van Straten and Bailes 2011). To eliminate the aliased (leakage) signals we then trimmed the spectral array down to 4096 channels, so the final bandwidth was approximately $3.56\,$MHz. With the exception of \S6.1 and \S6.2, all of the results presented in this section were obtained using the trimmed cyclic spectra from the second processing of our data.

\subsection{Bandpass Filter}
If we want to know the profile of the bandpass filter of our instrument there are two methods available to us: we can measure the average total power as a function of radio-frequency, or we can make use of the filter functions, $H$, obtained from our fitting. (One can also inject artificial pulsed power, of known spectral shape, into the signal chain, but we did not record such data.)

Our estimates of $H$ incorporate all of the filtering imposed on the signal. We expect there to be contributions from the ISM, the Solar wind, the Earth's ionosphere, and from our instrument (telescope, front-end and back-end). Of these various contributions, only the receiver system is expected to be stable over long time-scales. As $H$ is a complex quantity, averaging it will yield zero, but we can instead form $\langle|H(\nu)|\rangle$, which we take as an estimate of the bandpass filter, $|H_{rec}(\nu)|$.  Averaging over all filter solutions for all three epochs  we obtain the result shown in figure 1 for $|H_{rec}(\nu)|$. Also shown in figure 1 is the result of estimating the bandpass in a more traditional way, using the square-root of the average total power: $\sqrt{\langle D(0,\nu)\rangle}$. (The square-root appears here because the power is a quadratic function of the filter response.)

\begin{figure}
\figscaleone
\centerline{\plotone{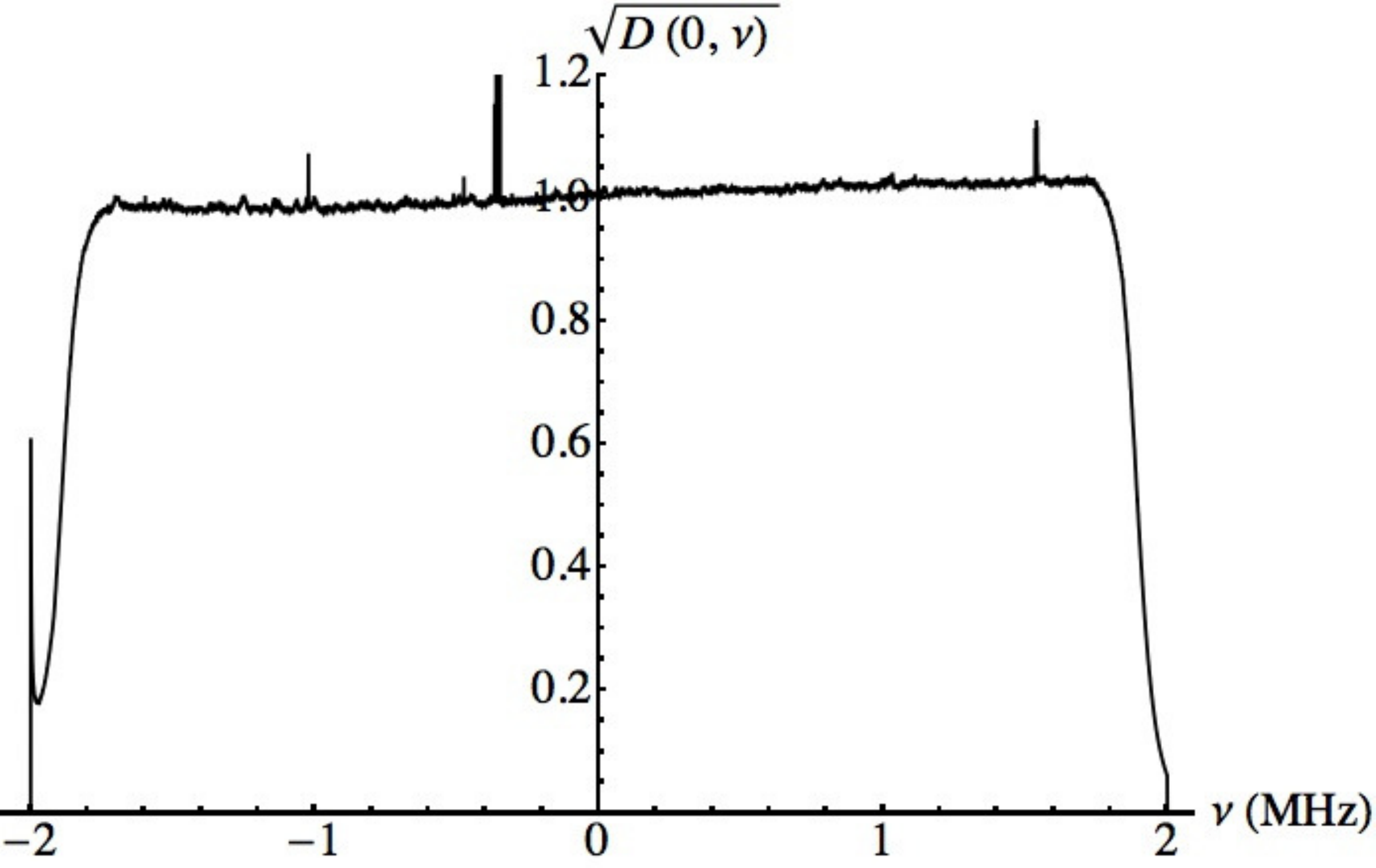}}
\hfill\break
\centerline{\plotone{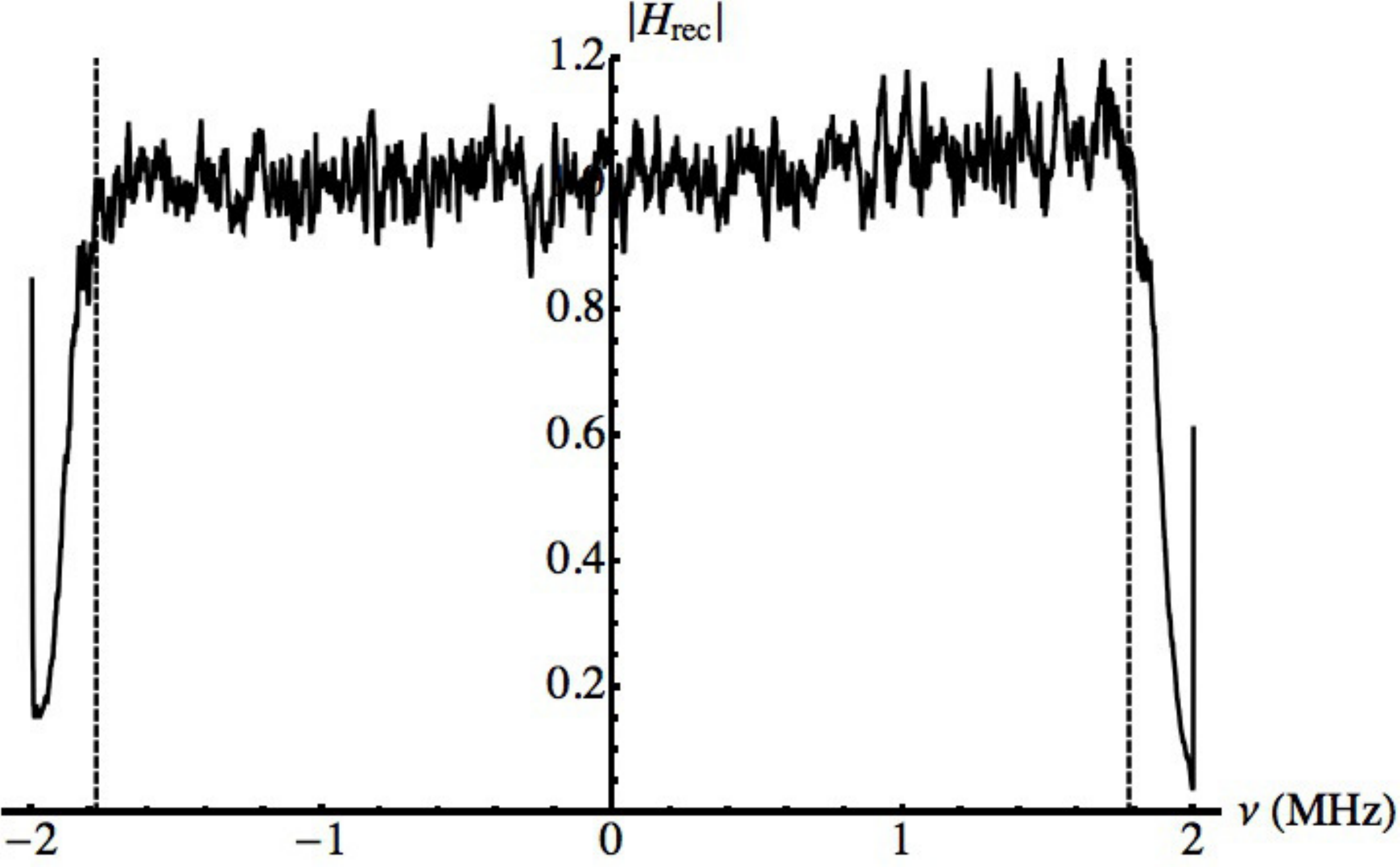}}
\caption{Two estimates of the amplitude of the instrumental bandpass filter for the ASP baseband recorder. The upper plot shows a traditional estimate for the bandpass, formed from the square-root of the total power $\sqrt{\langle D(0,\nu)\rangle}$, averaged over the data taken on MJD53791. The lower plot shows the result of averaging $|H(\nu)|$ over all three epochs of observation. The vertical, dashed lines in the lower plot delimit the regions which we trimmed off, to eliminate aliased signals leaking in at the edges of the band.}
\end{figure}

Although $H_{ISM}$ fluctuates quite rapidly, the amplitude of those fluctuations is large, so a long total observation time is required in order to form an accurate estimate of $|H_{rec}(\nu)|$. With our three epochs combined we have approximately 4.5~hours of data, and the scintillation time-scale is of order a minute so we expect our estimate of the filter response to be accurate to $\sim6$\%. That is approximately the level of fluctuation seen in our estimate of $|H_{rec}(\nu)|$ across most of the band. Thus the only clearly significant structure we find in $|H_{rec}(\nu)|$ is the roll-off of the filter at the band edges. A cause for concern is the abrupt rise in the estimated filter response at both extremes of the frequency range. These upturns indicate that that there is some leakage of signal from outside the nominal band of the filter. 

By contrast with $|H_{rec}(\nu)|$, the estimate $\sqrt{\langle D(0,\nu)\rangle}$ shows evidence of an upturn only at one end of the band. The reason for this difference is unclear. The other points of distinction between the two results are (i) that the noise on the traditional estimate is much smaller, even though only a third as much data was used, and (ii) Radio Frequency Interference (RFI) is manifest in the traditional estimate. To some extent the effect of the RFI could be mitigated by averaging using the median estimator, rather than the mean, but this would not help for steady interference. The reason for the lower noise-level on the traditional bandpass estimate can be seen from equation (10). Our solutions for  $H(\nu)$ -- whence the $|H_{rec}(\nu)|$ estimate -- are based on the pulsed power, i.e. $\alpha\ne0$, whereas the zero-modulation-frequency data, $D(0,\nu)$, are dominated by the system noise, $N(\nu)$, which is both large and unpulsed. 

As mentioned at the start of \S6, the leakage at the band edges, most evident in the lower panel of figure 1, can introduce low-level artifacts into our filter or pulse-profile estimates. Consequently we decided to fully reprocess our data, trimming off the edges of the band as we did so. The results described in \S6.3 and later sections of this paper were obtained from the second processing in which the spectral band was trimmed.

\subsection{Bootstrap approach to the intrinsic profile}
Lacking prior knowledge of the intrinsic pulse profile we are obliged, as in \S3.3, to commence our modelling using the observed, scattered pulse profile as an approximation to the intrinsic profile. We then obtain our first model of the filter function, for each sample cyclic spectrum, by fitting to the data in the way described in \S\S4,5. The filters obtained in this way are then used to obtain a better estimate of the intrinsic pulse profile, and the whole process is iterated, obtaining better approximations to $S_x$, and the various $H$, on each pass through the data.

Once an accurate model of the intrinsic profile is obtained, other data-sets for the same pulsar taken with the same instrumental configuration can use that profile to obtain model filters in a single pass through the data. But new instrumental configurations -- e.g. different observing frequencies  -- may force a return to the bootstrap approach. 

Because it requires multiple passes through the data, a bootstrap can be slow. We can, however, speed things up to some degree because at the second and subsequent profile-iterations we already have available a set of model filters appropriate to each recorded cyclic spectrum. These filters can be used to initialize subsequent models prior to optimization. As our model of the intrinsic pulse profile approaches the true intrinsic profile we expect the model filters to change very little between successive iterations, so this procedure should accelerate the optimization substantially. This expectation was borne out in practice, as we now describe.

To enable a rapid approach to the intrinsic profile we initially used a subset of the data (roughly 20 minutes of observation) from one epoch (MJD53873), iterating several times on this subset, and then adding in the rest of the data from this epoch in order to improve the signal-to-noise ratio of our profile estimate.  For the first set of filter solutions, using Proximate initialization, we found that on average 289 {\it NLopt\/} steps were required to fit each cyclic spectrum in the data subset. Subsequently, using the filter models obtained at the previous iteration as our starting point, the number of {\it NLopt\/} steps declined to 222, 17 and 14 for the second, third and fourth iterations, respectively.\footnote{These step-counts refer to the first processing of the data.} The small decrease in the required number of steps between the third and fourth iterations, contrasting with the large decrease between the second and third iterations, suggested that we had reached the noise floor of the data subset, so for subsequent iterations we utilized all of the data from MJD53873 --- a total of approximately 2 hours.

For iteration five we needed to obtain the first filter solutions for the bulk of the data from this epoch, using Proximate initialization, which required on average $241$ {\it NLopt\/} steps per cyclic spectrum. But for all subsequent iterations we were able to initialize our models using the previous set of filter solutions. We found that iterations six and seven required only 12 and 3, respectively,$^8$ {\it NLopt\/} steps for each cyclic spectrum, indicating very rapid convergence of our estimate of the intrinsic pulse profile.

\begin{figure}
\figscaleone
\centerline{\plotone{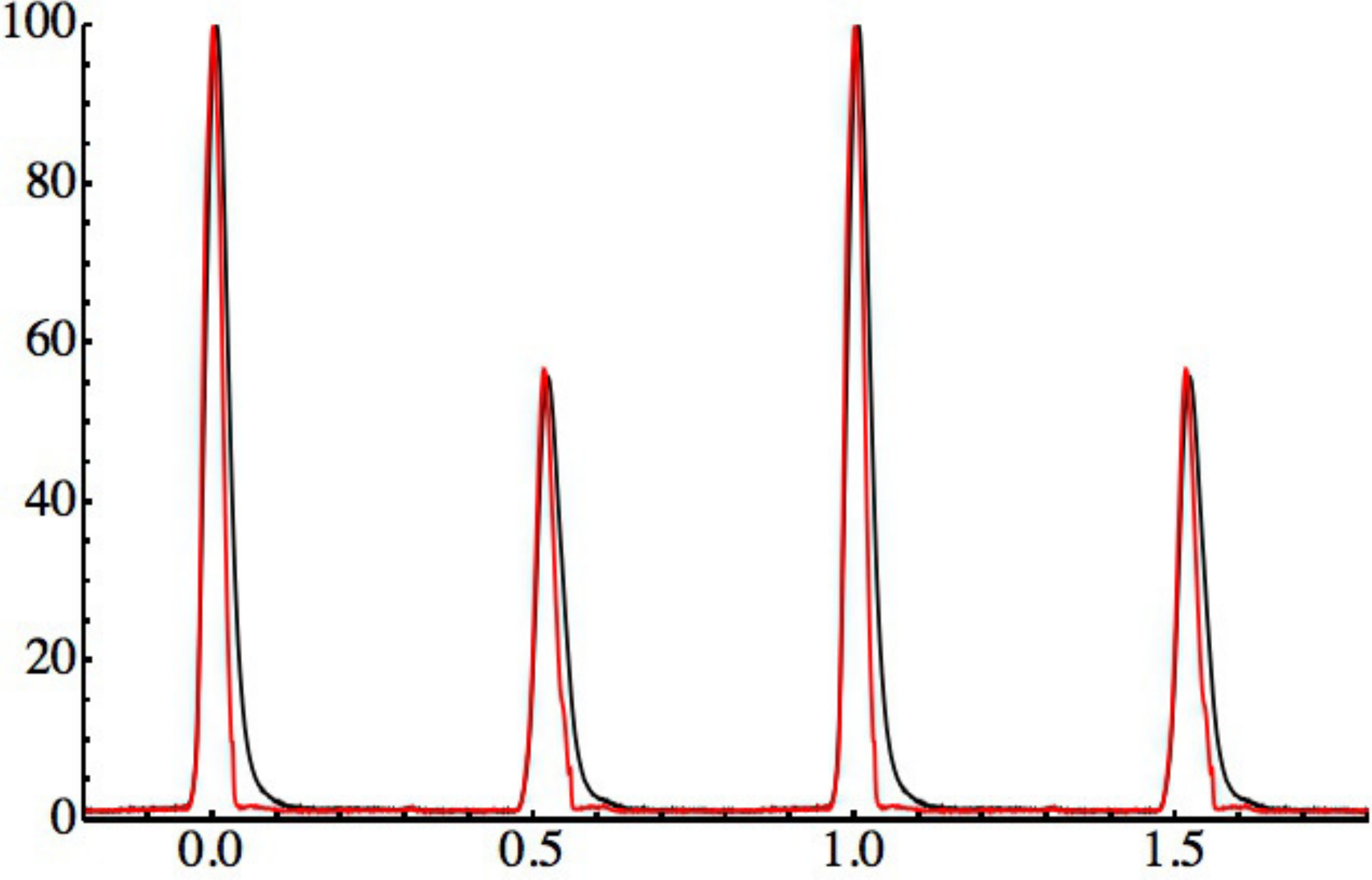}}
\centerline{\ \ \ \plotone{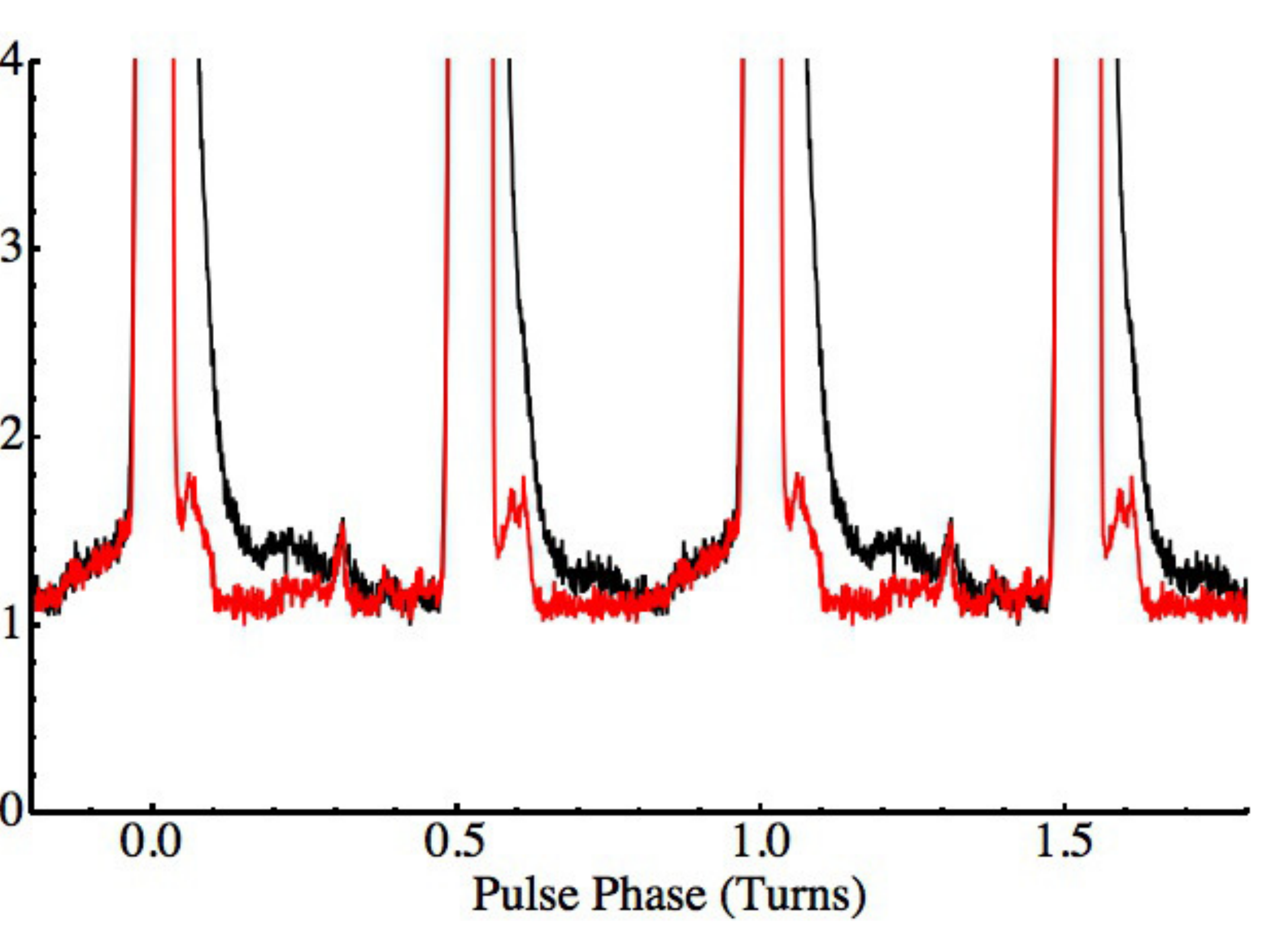}}
\caption{Intrinsic (red) and scattered (black) pulse profiles for B1937+21, at 428~MHz, observed on MJD53873. Two complete rotations are shown. The zero-point of the profile amplitude is arbitrarily chosen, whereas phase-zero corresponds to the peak of the (intrinsic) main-pulse. The top panel shows the full range of the pulse, while the lower panel shows a close-up of the lowest 3\%\ of the profile.}
\end{figure}

\begin{figure}
\figscaleone
\centerline{\plotone{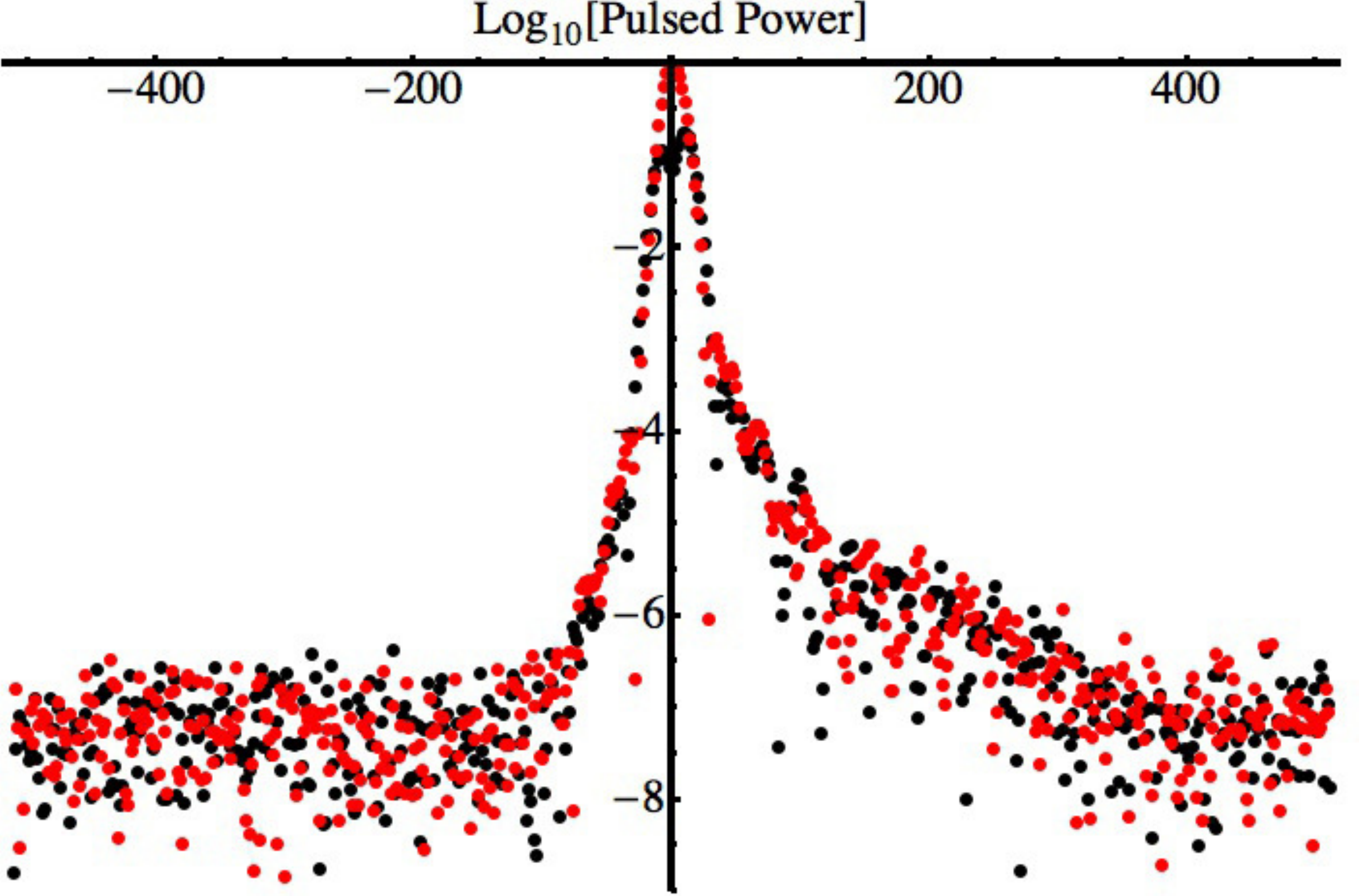}}
\caption{Intrinsic (right-hand-side: positive harmonics) and scattered (left-hand-side: negative harmonics) pulsed-power vs. harmonic number for B1937+21, at 428~MHz, observed on MJD53873. The pulse profile is real, so the power-spectrum is an even-function of the harmonic number. Odd-numbered harmonics (i.e. $\alpha=(2m+1)\Omega$, with $m$ an integer) are shown in black; even-numbered harmonics are shown in red.}
\end{figure}

\begin{figure}
\figscaleone
\centerline{\plotone{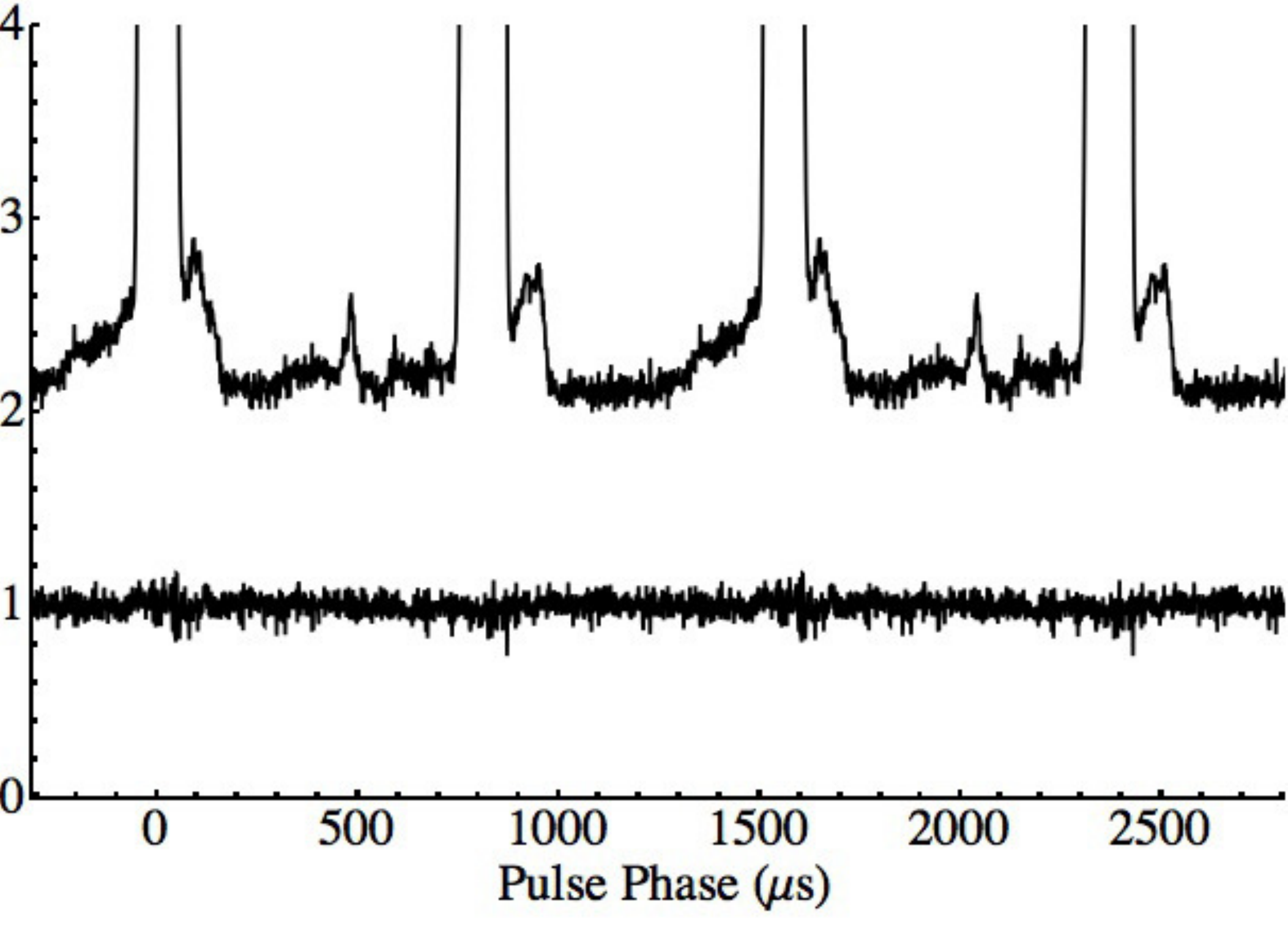}}
\caption{Comparison of intrinsic pulse profiles derived independently for MJD53873 and MJD53791. The mean of these profiles is shown in the upper curve, while the difference is shown in the lower curve. The scaling of this plot is as for figure 2, so the full scale of the mean profile has a range of 100. For clarity of presentation, arbitrary vertical offsets have been applied to both the mean and the difference profiles. Two complete rotations are shown.}
\end{figure}

\begin{figure}
\figscaleone
\centerline{\plotone{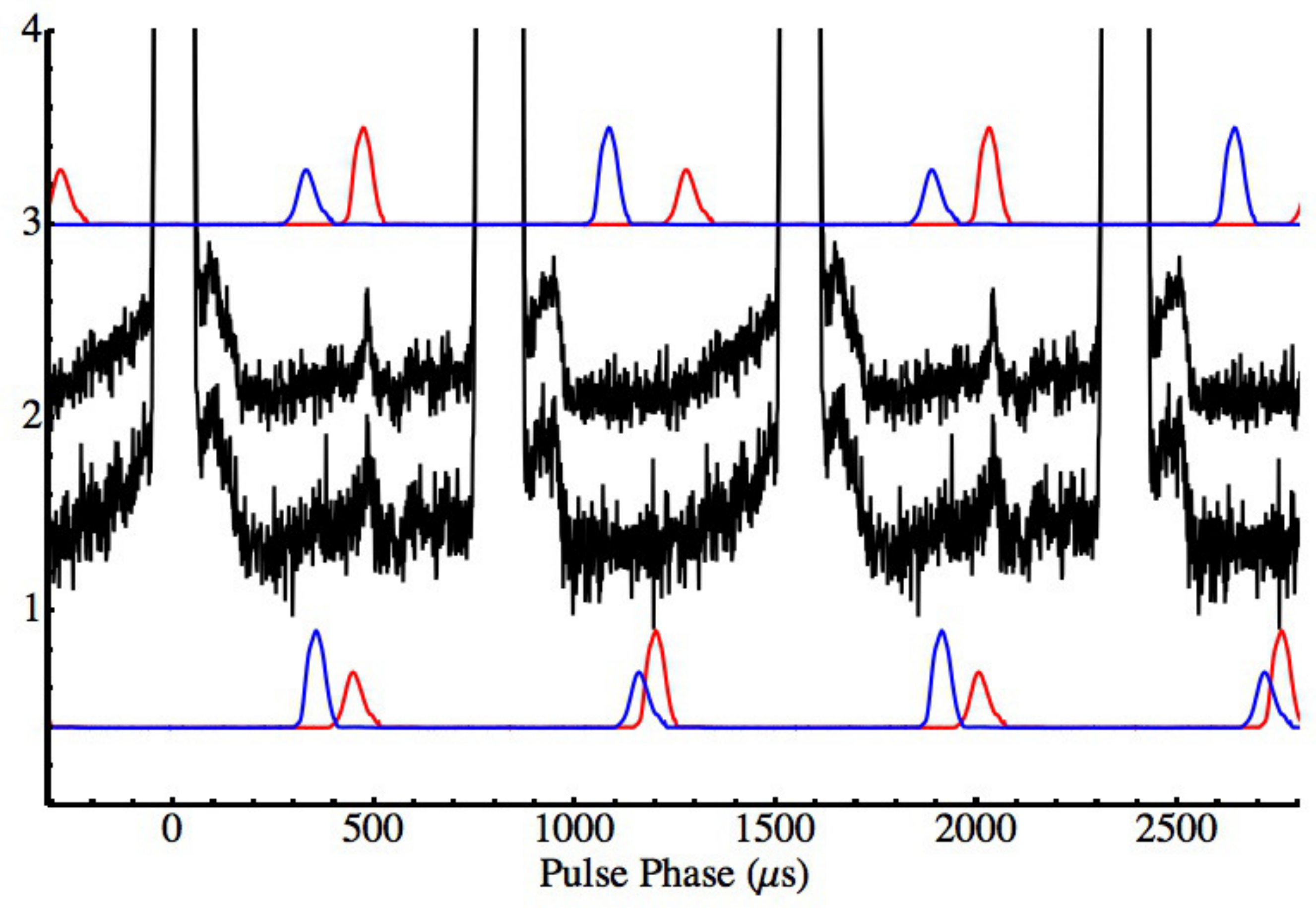}}
\caption{Comparison of intrinsic pulse profiles derived independently at 428~MHz (upper, black line) and 432~MHz (lower, black line) for MJD53791. The scaling of this plot is as for figure 1; arbitrary vertical offsets have been applied, for clarity, and two complete rotations are plotted. The red/blue curves show calculated profiles appropriate to leakage signals at the lower/upper edge of the 4~MHz band of the 428~MHz (upper curves) and 432~MHz (lower curves) data. These signals are delayed/advanced by roughly 30~ms, as a result of aliasing and the associated incorrect dedispersion. Normalisation of the red/blue curves is arbitrarily chosen. We find no indication of residual contamination by signal leakage in our results (see text, \S6.3).}
\end{figure}

Separately we have observed, when using an existing set of optimized filter models as our starting point, that our code requires a minimum of 3 {\it NLopt\/} steps to return an optimized solution, even when the same reference profile is used for both solutions. We therefore conclude that our intrinsic profile estimates for B1937+21 do not differ significantly between iterations six and seven, and further iterations are unwarranted.

Use of the previous set of filter solutions to initialize our models clearly leads to a substantial saving in computation time. Using Proximate initialization we expect that the bootstrap would have required a total of 10~days of CPU time, whereas the sequence just described used only a third of that time. In fact our procedure needed only one quarter more time than a single pass through the same data using a given reference pulse profile.

\bigskip

\bigskip

\subsection{Intrinsic versus scattered profile}
In figure 2 we show our estimate of the intrinsic profile, together with the scattered profile, using all the data from MJD53873. This epoch was chosen because we obtained significantly more data on that date than on either of the other epochs. As expected, the intrinsic modulation profile of the signal is much sharper than the apparent modulation, because of the contribution of the scattered (delayed) waves to the apparent profile. The ``scattered tail'' of the pulse is absent from our estimate of the intrinsic profile.

Figure 2 (lower panel) also reveals the presence of several low-level (a fraction of 1\%\ of the peak height), but sharp features in the ``baseline'' of the intrinsic pulse. These features are difficult to recognise in the scattered profile for two reasons. First, interstellar scattering broadens them, while decreasing the peak amplitude of each. Secondly, the features that are present immediately after the main-pulse or the inter-pulse are swamped by the delayed signal from those two, very strong components of the pulse profile.

An equivalent description of the pulse modulation is available by Fourier-transforming the scattered and intrinsic profiles. The resulting harmonic powers are shown in figure 3, demonstrating that the high harmonics of the intrinsic profile contain a great deal more power than the scattered profile. This is just as expected. The scattered profile is a convolution of the intrinsic profile with the impulse response function,  so in the Fourier domain the relationship is multiplicative, and the multiplier declines from near unity at low harmonic numbers to very small values at high harmonic numbers.

Because the low-level features evident in figure 2 are seen here for the first time at these radio frequencies, and the signal-processing we have used to reveal these structures is itself novel, we would like to have some confirmation of their reality.  We have therefore undertaken a completely independent bootstrap estimate of the intrinsic profile for another epoch, MJD53791. In this comparison we are not interested in any timing (pulse-phase) offset between the two epochs, so in comparing the two intrinsic profiles we have applied a pulse-phase shift and a scaling, chosen so as to minimize the difference between the profiles.

The result of our two independent bootstrap solutions can be seen in figure 4, where we show the mean of the intrinsic profiles and their difference. The latter curve appears noise-like, without any clearly significant differences between the two, independently derived intrinsic profiles. In particular we note that the largest differences occur underneath the main-pulse and inter-pulse components, where the signal is very strong and the noise is therefore greater than at other pulse-phases. There is no apparent systematic difference between the two profiles at those pulse-phases where the weak, low-level features are seen.

As a final check on the reality of the features revealed in figures 2, 4, we have also compared the intrinsic pulse profiles obtained from independent bootstrap estimates at two different frequencies, 428~MHz and 432~MHz, for the epoch MJD53791 --- this comparison is shown in figure 5. Although the 432~MHz data exhibit more system noise than the 428~MHz profile, because the integration time for the latter is larger by a factor of 1.5, the two profiles appear otherwise very similar in respect of the low-level features which are revealed by construction of the intrinsic profile. 

An important aspect of the inter-band comparison in figure 5 is that it excludes signal leakage (\S6.1) as a possible origin for the low-level structures which we see in the intrinsic profile. Even though we have trimmed the band edges, which should eliminate the bulk of that problem, it is possible that some traces of leakage remain.  This concern is heightened by the fact that the sharp feature at a pulse-phase of $500\,{\rm\mu s}$ lies close to the expected location of the aliased main pulse component, for signals leaking across the low-frequency edge of the 428~MHz band (upper red curve in figure 5).  The inter-band comparison makes it plain that this is not a viable explanation for that feature, because at 432~MHz the corresponding alias should lie at 1,200$\,{\rm\mu s}$, where no profile feature is seen -- yet the observed $500\,{\rm\mu s}$ peak appears very similar in the two bands. We also note that interpreting the $500\,{\rm\mu s}$ feature as an alias of the main-pulse implies that there should be a counterpart feature from the inter-pulse, roughly half a turn later, whereas no such feature is observed in either band. Overall, the aliased signals from the band edges do not correspond with the low-level features we see in the pulse profiles in either band, and we conclude that they are not due to out-of-band signals. 

In fact residual out-of-band signals are expected to appear as broad structures in the time-domain, because the dispersive delay is a strong function of frequency. The sharpness of the features shown in the red and blue curves in figure 5 is due to the fact that only frequencies immediately adjacent to the band edges have been considered. The red and blue curves are simply calculated as delayed (and scaled) versions of the mean pulse profile, with the delay/advance equal to the difference in dispersive delay between the upper and lower edges of the band. At MJD53791 the dispersion measure of PSR B1937+21 was 71.023$\;{\rm pc\,cm^{-3}}$, and the period was 1.5577~ms, so the aliased signals appear at $\pm30.068\;$ms (428~MHz band) and $\pm29.240\;$ms (432~MHz band). Modulo the pulse period these become, respectively, $\pm0.470$ and $\pm1.201$ milliseconds.

Some of the ``new'' structure that we see in the intrinsic pulse profile corresponds well with features of B1937+21 which have been found by others, as follows. The distinct peaks seen immediately after the main- and inter-pulse have previously been observed by a number of authors at higher radio-frequencies, where the delayed, scattered signal is much weaker --- see, particularly, figure 1 of Thorsett and Stinebring (1990). Here we are presumably seeing the emission regions which are responsible for the giant pulses of B1937+21 (Cognard et al 1996), and the consequently high modulation index at these pulse-phases (Jenet, Anderson and Prince 2001). The sharp feature we see at a pulse-phase of $500\,{\rm\mu s}$ ($0.3$ turns) has a counterpart which was noted in L-band observations by Yan et al (2011). Residual dispersion smearing in the Yan et al (2011) data is significant, so it is not surprising that their feature appears broader than the one we observe. Finally, the gradual rise we see in the $0.2\;$turns immediately preceding the main-pulse is also manifest in the Yan et al (2011) data. 

\medskip

The consistency of our intrinsic profile estimates across different epochs and spectral sub-bands, and the connections we can make between individual features and previous observations of B1937+21 at other frequencies, give confidence that the statistically-significant features we see in our intrinsic profile are indeed real.

\begin{figure}
\figscaleone
\centerline{\plotone{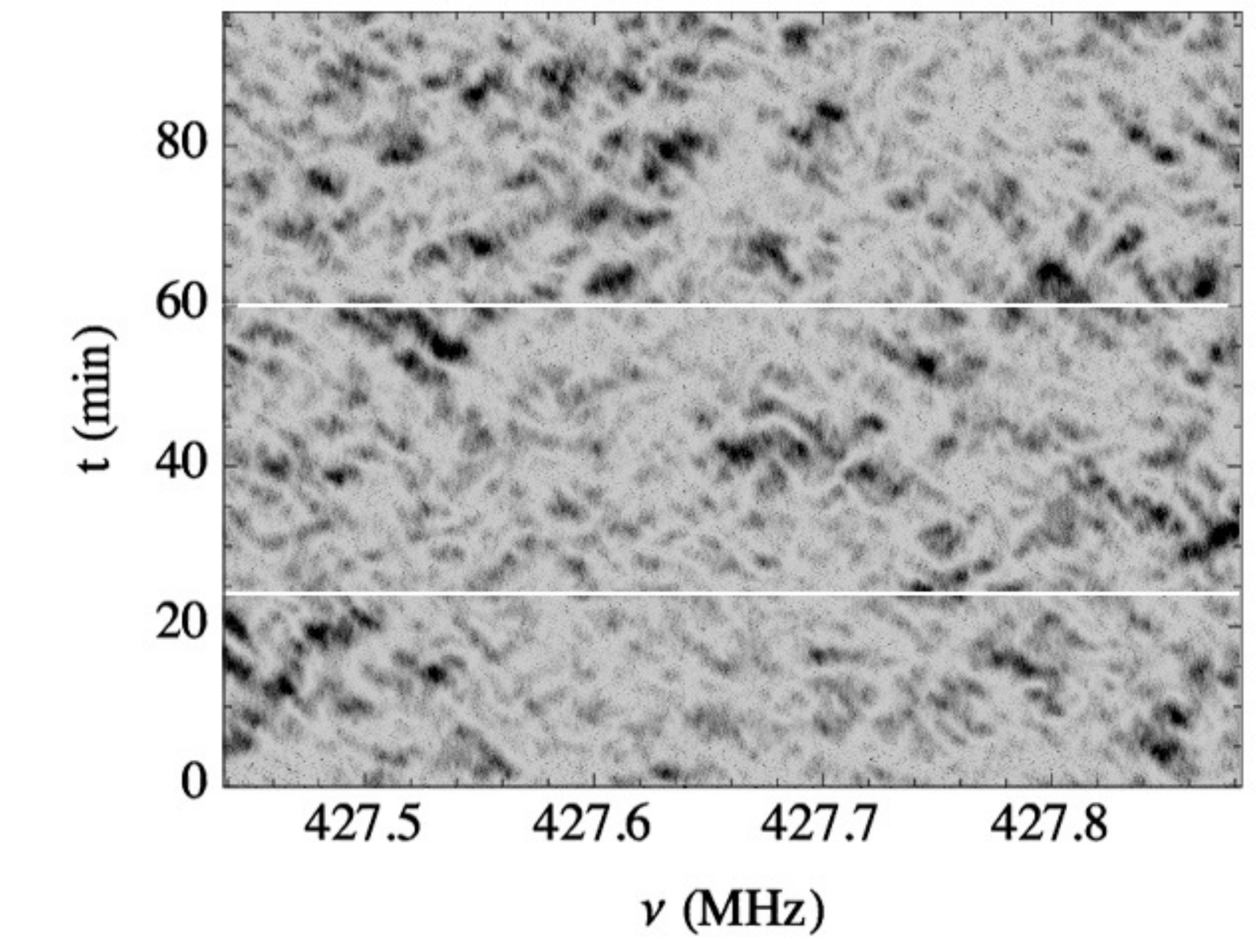}}
\hfill\break
\centerline{\plotone{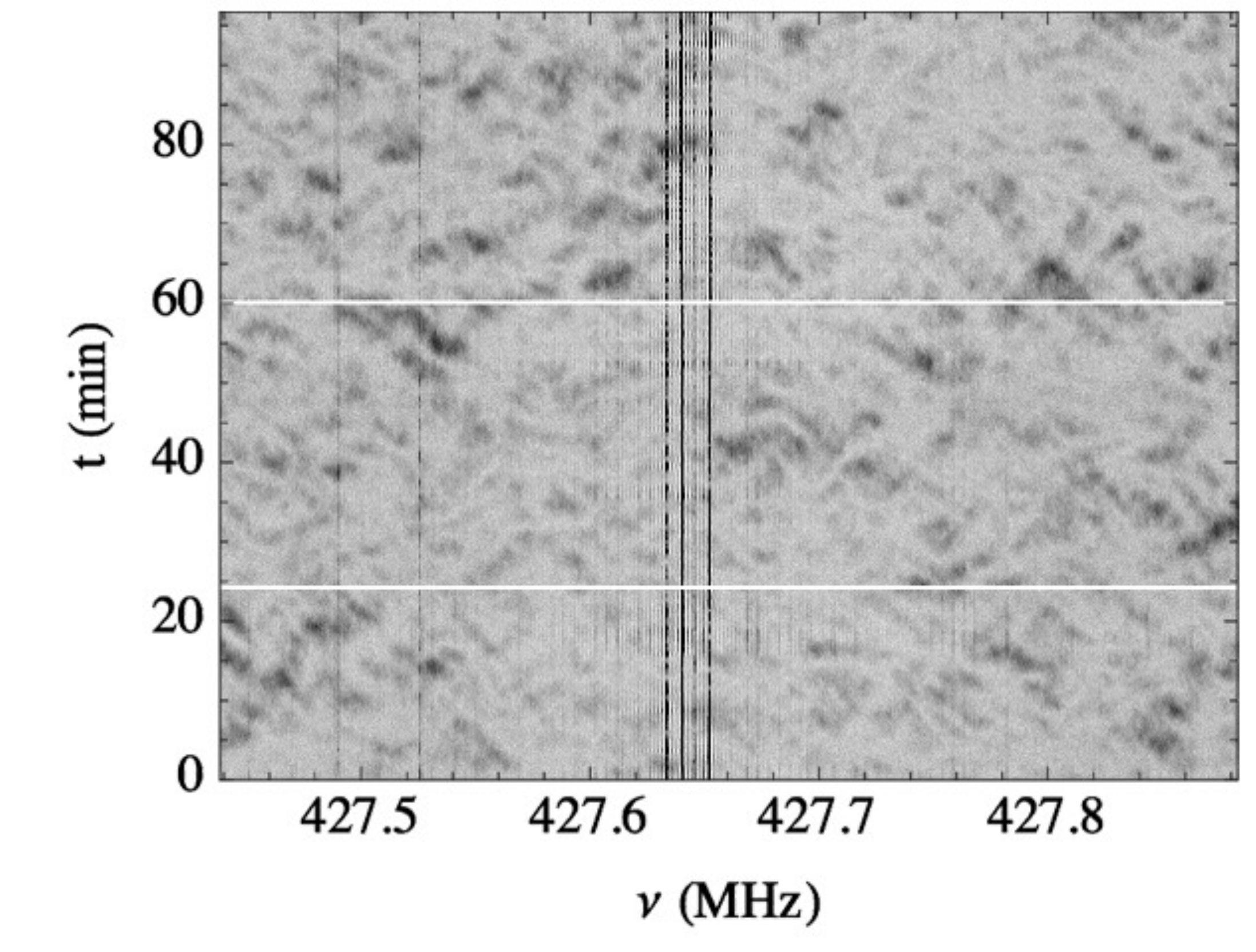}}
\caption{Inverted grey-scale image of the dynamic spectrum, $D(0,\nu,t)$ (lower panel), recorded on MJD53791, together with the corresponding dynamic filter power, $|H(\nu,t)|^2$ (top panel). Only a fraction ($\simeq0.44\;$MHz) of the recorded bandwidth is shown; the temporal extent is approximately 98~mins. Two short gaps in the temporal coverage are visible as discontinuities, running horizontally in both images. Radio-Frequency Interference is evident in the dynamic spectrum as thin, black, vertical lines; but it is almost completely absent from the dynamic filter power.}
\end{figure}

\subsection{Dynamic Spectra}
A measured cyclic spectrum quantifies the power spectrum of the signal as the zero-modulation-frequency array $D(0,\nu)$ (see \S2). We compute our cyclic spectra with a cadence of 15 seconds, and thus we can trivially obtain a dynamic spectrum from the temporal sequence of $D(0,\nu)$. This dynamic spectrum is a simple time-average, not a difference of on-pulse and off-pulse power, so it includes all power contributions: noise from the receiver and the sky, the pulsar signal, and any terrestrial signals reaching the receiver, i.e. RFI. Because RFI can cause severe problems for some types of radio astronomical investigations, it is useful to examine the dynamic spectrum in order to gauge its impact.

Figure 6 shows the dynamic spectrum for a 512-channel spectral segment recorded on MJD53791; RFI is manifest in this segment as narrow spectral lines. None of these lines is so strong that the voltage signal exceeds the dynamic range of the sampler, nor is any impulsive RFI evident in figure 6. These aspects of the data reassure us that the observations were taken under relatively benign RFI conditions, and in this circumstance we can reasonably expect a high level of immunity from RFI in our models of $S_x$ and $H$. In particular, if the RFI is both accurately captured and not modulated at the frequency $\Omega=1/P$, or its harmonics, then cyclic spectra will be free of RFI contamination. 

To demonstrate that the observed RFI does not propagate into our model filters we also show in figure 6 the squared-modulus of the dynamic filter, i.e. $|H(\nu,t)|^2$. This quantity is our estimate of the contribution of the pulsar to the dynamic spectrum; the spectral structure $|H(\nu,t)|^2$ can also be seen in the total power signal. It is evident that the RFI present in the total power signal is absent from the dynamic filter.  We emphasise that the specific, small fraction of the spectrum shown in figure 6 was chosen at random: it was {\it not\/} selected because it displays good immunity from RFI.

\begin{figure}
\figscaleone
\centerline{\plotone{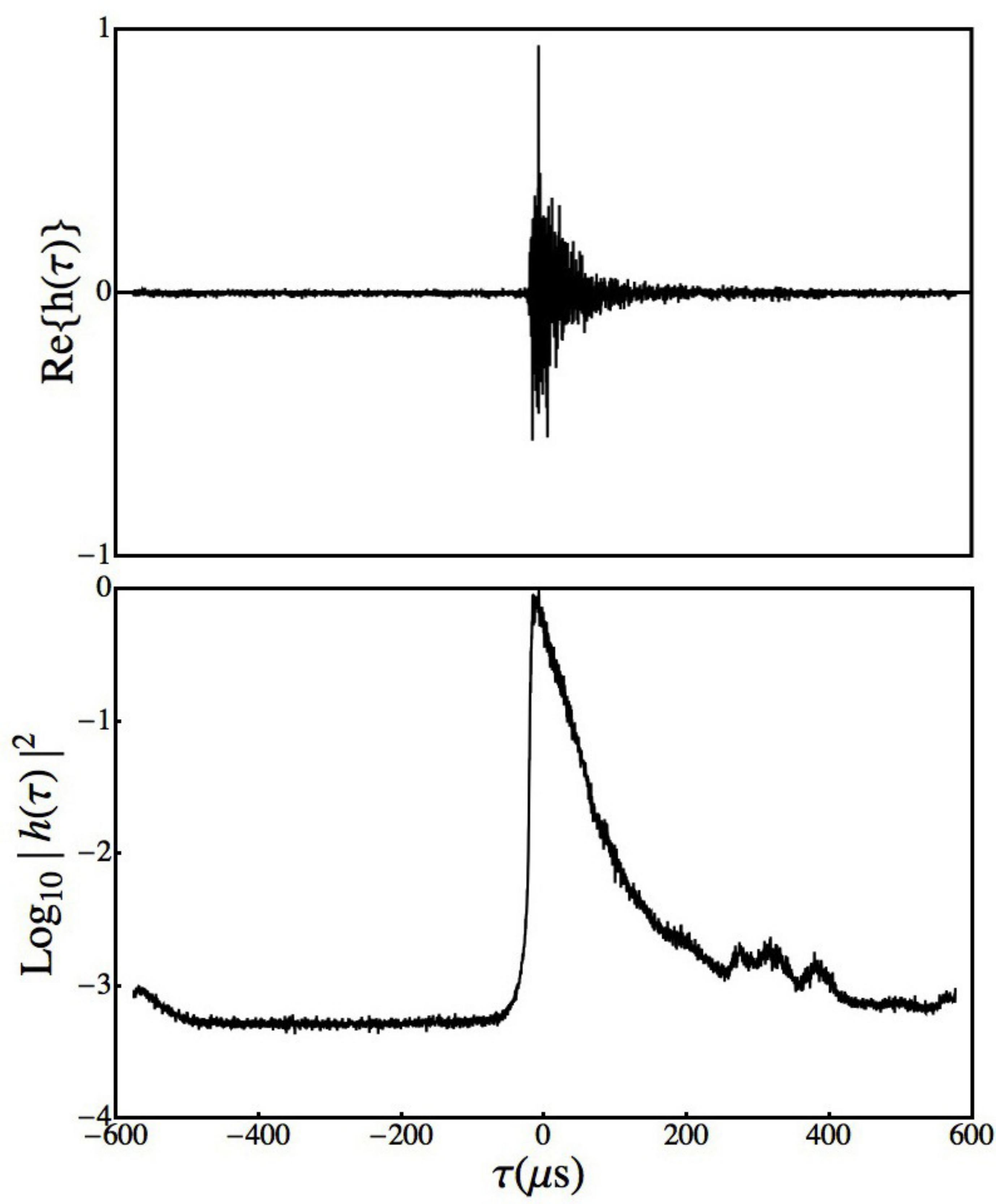}}
\caption{The impulse response function, $h(\tau)$, for B1937+21 observed on MJD53873. The top panel shows the real part of $h$ (linear scale, arbitrary normalization) for the first cyclic spectrum recorded at that epoch, while the lower panel shows the average $\langle|h(\tau)|^2\rangle$ (normalized by the maximum of $\langle|h(\tau)|^2\rangle$) over all the data taken at that epoch. The data in both plots covers a total range of  1,152$\,\mu$s in delay. The impulse response function itself is characterized by 4,096 complex coefficients, evenly spaced in lag.}
\end{figure}

\subsection{Dynamic fields}
Whereas the dynamic spectrum is a quantity which pulsar astronomers routinely measure, it has been much more difficult to get at the dynamic electric field because the latter requires information on the phases, and that information is usually not explicit in the measured intensities. The requisite phases can sometimes be retrieved -- e.g. if the field is sparse in some representation -- but to date this has been successfully demonstrated for only one dynamic spectrum (Walker et al 2008). By contrast, cyclic spectroscopy provides us with access to the electric field envelope, including the phase information; as such it is an intrinsically holographic method.

There are various possible representations of the dynamic fields because they may be described in terms of frequency-space (filter) or lag-space (impulse-response) coefficients, and the dynamic nature of the field can be represented either as a temporal sequence or in terms of the conjugate Fourier variable, i.e. a frequency.

\subsubsection{Impulse-response functions}
Figure 7 (top panel) shows one possible representation of the field: the real part of the impulse-response function, $h(\tau)$, determined from the first cyclic spectrum we observed on MJD53873. This function spans a lag range of $1,\!152\,{\rm\mu s}$, and we see that the amplitude of the response falls off on lag-scale $\la 50\;{\rm\mu s}$. There is, however, a low-level tail to the response, extending to lags that are a substantial fraction of the pulse period. To bring out these low-level features we took the modulus of the impulse response, and then averaged it over all the data at this epoch of observation. The lower panel of figure 7 presents the resulting $\langle|h(\tau)|^2\rangle$, which demonstrates that the low-level tail of $h$ continues out to delays of at least $400\;{\rm\mu s}$ relative to the peak of the response.

At extreme negative lags there is an obvious rise in $|h|$. The origin of this feature is not completely clear; however, a preliminary analysis suggests that parameter covariances in $\{h_j\}$ (see Appendix) may give rise to increased noise near the lag limits of the cyclic spectra, and we therefore consider this to be an artifact. 

On the other hand the features seen in the vicinity of $\tau\sim+300\;{\rm\mu s}$ appear to be bona fide structure in $h$. The delay-Doppler image, which we present in the next section, gives more information on these features.

\subsubsection{Delay-Doppler field images}
Finally we present our results in the Fourier domain conjugate to $(\nu,t)$. The conjugate variables $(\tau,\omega)$ have immediate physical meaning as the delay and Doppler-shift, respectively, that accumulate during propagation of the wave (Harmon and Coles 1983; Cordes et al 2006).  The Fourier Transform, $h(\tau,\omega)$, of the dynamic electric field,  $H(\nu,t)$, is therefore a quantity of particular interest we call this the ``delay-Doppler image''. Figure 8 shows the squared-amplitude of the delay-Doppler image for our data taken on MJD53873.

\begin{figure}
\figscaleone
\centerline{\plotone{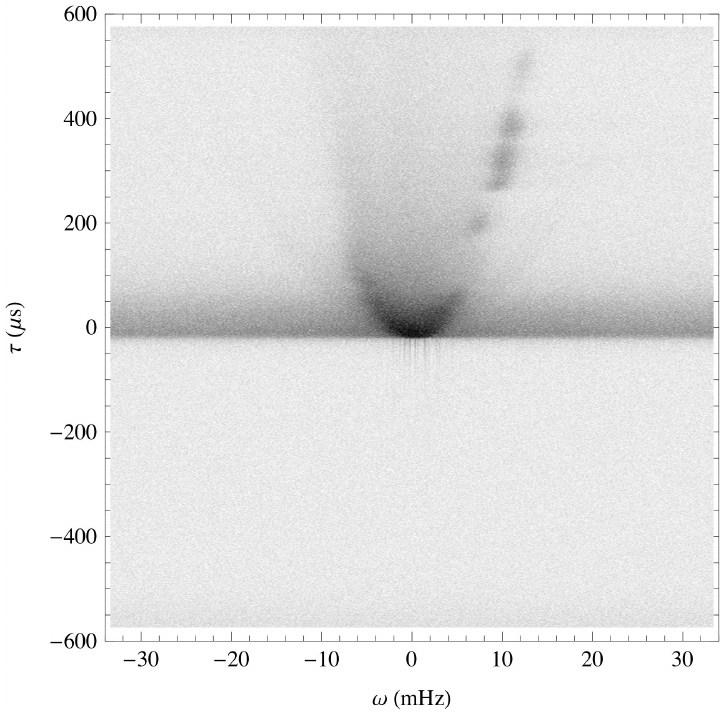}}
\caption{The squared-modulus of the delay-Doppler field image, $|h(\tau,\omega)|^2$, for B1937+21 observed on MJD53873. The field intensity is represented as an inverse, logarithmic grey-scale, over a range of 50~dB (almost the full dynamic range of the image, which is 51~dB). In this image the vertical dimension is delay, spanning the range $|\tau|\le576\,{\rm\mu s}$, and the horizontal dimension is Doppler-shift, with $|\omega|\le100/3\,{\rm mHz}$.}
\end{figure}

\begin{figure}
\figscaleone
\centerline{\plotone{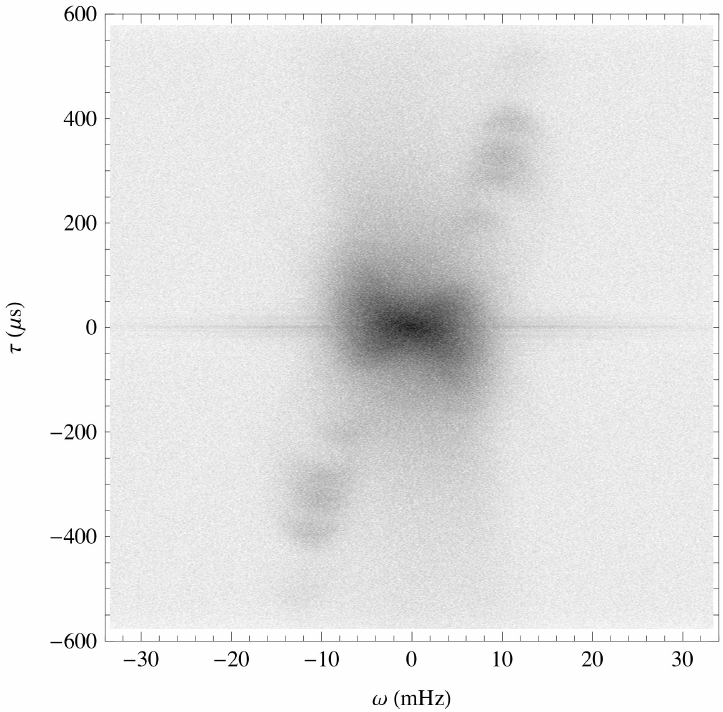}}
\caption{The power-spectrum of the dynamic-spectrum (often called the ``secondary spectrum''), $|{\cal F}\{|H(\nu,t)|^2\}|^2$, for B1937+21 observed on MJD53873. Power is represented as an inverse, logarithmic grey-scale, over a range of 50~dB. The full dynamic range of the secondary spectrum -- i.e. the ratio of the peak value to the noise-floor -- is 80~dB.}
\end{figure}

The lower-half of figure 8 is largely free of signal, as expected for negative lags (which are acausal). The only signals that can be recognised at negative lags are the band of scattered power running horizontally across the figure (discussed later), and a handul of thin, faint, vertical streaks in the region $|\omega|\la4\;$mHz, $0>\tau\ga-100\;{\rm\mu s}$.  We are uncertain as to the cause of these streaks, but we suspect that they may be sidelobes caused by the sharp truncation of the spectrum which we introduced by trimming the band (\S6.1). These streaks were not seen in the delay-Doppler image that we obtained in our first processing of the data. The enhanced noise at extreme negative lags, plainly seen in the average signal in figure 7, is also present in figure 8 but is difficult to discern without averaging.

By contrast, in the upper half-plane of figure 8 there is an abundance of structure. Most of the power is concentrated in a broad distribution centered on zero-Doppler-shift. And the overall distribution appears to have an approximately parabolic envelope, as is now familiar for many pulsars (Stinebring 2001; Cordes et al 2006). But there are also some discrete concentrations of power. Most evident of these are the concentrations in the range ${\rm 200\la\tau(\mu s)\la400}$ on the right-hand-side of the figure. These concentrations indicate that there are particular regions, within a few milli-arseconds of the direct line-of-sight to B1937+21, which are strongly diffracting, or refracting signals from this pulsar into our telescope. Apparently similar features were discovered by Hill et al (2005), in a multi-epoch study of PSR B0834+06, who found that their features appear to move through the delay-Doppler plane at constant velocity, consistent with the observed proper-motion of the pulsar. At present we don't know whether that property also holds for the features seen in figure 8.

In addition to the real structure just discussed, a strong artifact is plain in figure 8: around zero delay there is a broad, horizontal stripe in the image. The nature of this feature is clear: it is ``scattered power'' caused by discontinuities between successive values of $H(\nu)$ (or $h(\tau)$) in our temporal sequence. These discontinuities might arise in several ways, for example: inadequate sampling of the evolving $H(\nu)$; amplitude fluctuations in the pulsar; arbitrary phase rotations between successive filter solutions (per the degeneracy in overall phase, \S2.1); or gaps in the data record. We have considered each of the above possibilities, but none provides a satisfactory explanation, as we now detail.

First, the evolution of the filter $H(\nu)$ is well sampled by our 15-second cadence, as can be seen from the upper panel of figure 6. Secondly, there are $\sim10^4$ pulses within each of our cyclic spectra, so the variations in average intensity between samples will be small, $\la1$\%. In fact even this variation is irrelevant to figure 8 as we have normalized each filter solution such that it has a root-mean-square value of unity. Thirdly, the arbitrary phase of each filter (see \S2.1) has been chosen so that each solution $H(\nu,t_n)$ matches the previous solution $H(\nu,t_{n-1})$ as closely as possible, in a least-squares sense. Finally, although there is indeed a gap of 30~seconds in our temporal coverage (caused by a change of hard-disk during observing), we have interpolated across this gap before constructing figure 8. For these reasons we do not expect any of these effects to be responsible for the high levels of scattered power seen in figure 8.

A clue to the origin of the scattered power can be found in figure 9, which shows the ``secondary spectrum'' -- i.e. the power-spectrum of the dynamic spectrum, $|H(\nu,t)|^2$ -- for our data. By contrast with figure 8, this quantity shows quite low levels of power scattered to large Doppler-shifts. In forming the spectrum, $|H(\nu)|^2$, we are erasing all information on the phase of the filter $H(\nu)$, so the difference in scattered power levels between figures 8 and 9 indicates that the source of the scattered power in figure 8 is phase-discontinuities between adjacent filters, $H(\nu)$. As noted above, we have matched the phases of adjacent filters, to the extent that this can be done with a uniform phase rotation of $H$. Therefore our filter solutions contain non-uniform phase structure that is discontinuous between adjacent samples. In \S4.3.1 we noted that our filter solutions may exhibit covariance between the lag coefficients $h(\tau_m)$ and $h(\tau_j)$, for lag separations small compared to the pulse-width ($|\tau_m-\tau_j|\ll w$), and that the poorly constrained combination ($h(\tau_j)-h^*(-\tau_j)$) modifies only the phase of $H$. We therefore attribute the scattered power evident in figure 8 to these parameter covariances. We defer a detailed treatment of these issues to a later paper.

As figures 8 and 9 both display the response of the interstellar medium in the delay-Doppler coordinate frame, it is worth clarifying the relationship between them. Recall that $h(\tau,\omega)$ is just the Fourier Transform of the sequence $H(\nu,t)$. Thus the Fourier transform of the dynamic spectrum, which is the Fourier transform of the product $H(\nu,t)H^*(\nu,t)$, is just the convolution of $h(\tau,\omega)$ with $h^*(\tau,\omega)$. Consequently the arc that appears around the origin in figure 8 is echoed in a series of inverted arclets in figure 9; each of these arclets is centered on one of the power concentrations visible in figure 8 --- cf. figure 5 of Walker et al (2004). Because the ``secondary spectrum'' (figure 9) is equivalent to a self-convolution of the delay-Doppler image (figure 8), the latter is more fundamental and will typically be the more useful quantity for two reasons. First because the delay-Doppler image exhibits the scattered field with greater clarity: in the secondary spectrum the scattered field is tangled up with itself. Secondly, convolution is a smoothing operation, so faint power concentrations are more easily seen in the delay-Doppler image. These points are well demonstrated by comparing figures 8 and 9.

Despite the fact that figure 9 is derived from the dynamic spectrum, it could not have been obtained by conventional spectroscopic methods, in which the on-pulse power-spectrum is determined within a window of width comparable to the width of the main-pulse (or inter-pulse) component. The reason is simply that windowing restricts the lag-range of the resulting secondary spectrum to the width of that window. Refering to figure 2 we see that the main-pulse would be completely contained within a window of width $\sim100\,{\rm\mu s}$, so the resulting lag range would be $-50\le\tau({\rm\mu s})\le50$ --- a tiny fraction of the actual lag range of figure 9. 

\section{Discussion and future directions}
Because cyclic spectroscopy has not previously been applied to radio pulsar signals, there are many related issues that deserve consideration. Here we confine ourselves to a brief discussion of three aspects that the present study calls attention to.

\subsection{Precision timing of PSR B1937+21}
It is well known that the small-scale structure of the ISM can have a significant effect on the measured arrival times of radio pulses, in consequence of the delays (geometric and wave-speed) associated with signal propagation (e.g. Foster and Cordes 1990). These effects are of particular importance if they are epoch dependent, which is the case if the scattering properties of the medium are not statistically uniform transverse to the line-of-sight. It is plain from figure 8 that some of the scattering material towards B1937+21 is indeed very clumpy, with several flux concentrations appearing far from the origin, albeit at low power levels. Previous studies of the dispersion and scattering on this line-of-sight (Cordes et al 1990; Ramachandran et al 2006) preferred a near-Kolmogorov model of the structure, but the clumpiness we see is quite different from the expectations of a uniform Kolmogorov model (Cordes et al 2006; Walker et al 2004). As B1937+21 is routinely used for precision timing experiments (e.g Verbiest et al 2009), a better understanding of this scattering material is desirable.

It has previously been reported (Cognard et al 1993; Lestrade, Rickett and Cognard 1998) that B1937+21 occasionally exhibits timing fluctuations, correlated with flux variations, whose properties are suggestive of  ``Extreme Scattering Events'' --- that is, plasma-lensing events (Fiedler et al 1987, 1994; Romani, Blandford and Cordes 1987). Such events require close alignment between the observer, plasma-lens and pulsar, and these events are consequently rare. If the alignment is not so close then the lens will cause smaller flux changes, but may still have a significant effect on the pulse arrival time because the extra path-length traversed by the faint images may be large. Furthermore these poorly aligned lens configurations should be relatively common. It is possible that plasma-lensing is responsible for the discrete flux concentrations that we see in the vicinity of $\tau\sim300\;{\rm\mu s}$ (figure 7 and 8), with each concentration being due to one or more additional faint images. We note that at this epoch (MJD53873) the features appear at such large delays that the scattered pulse has little overlap with the unscattered signal, so the pulse arrival time estimate should not be greatly affected. But at later epochs, when the scattering structures are closer to the line-of-sight to the pulsar, the scattered signals may appear at delays $\tau\sim100\;{\rm \mu s}$ where they can exert a substantial influence on the measured arrival time. We defer a quantitative examination of pulse arrival time variations to a later paper.

Depending on the electron column-density structure, and the pulsar-lens-observer configuration, several additional images may arise from one plasma lens, so it is possible that all of the flux concentrations we see near $\tau\sim300\;{\rm\mu s}$ in figures 7 and 8 are due to a single lens. Under that hypothesis, the observed range of delays ($200\la\tau({\rm\mu s})\la400$) tells us something about the size of the lens. Assuming that the pulsar is at a distance $\sim5$~kpc, and that the lens is near the midpoint, one finds that the lens diameter is $\sim4\;$AU. This is comparable to the dimensions that have previously been inferred for the lenses responsible for Extreme Scattering Events (e.g. Romani, Blandford and Cordes 1987).

Unfortunately, with the techniques currently available to us, it is not possible to distinguish between lens-like, refractive behaviour and diffractive scattering as the cause of the observed power concentrations around $\tau\sim300\;{\rm\mu s}$. The clearest way to distinguish between these possibilities would be to undertake rigorous, quantitative physical modelling of the particular wave-propagation paths for this line-of-sight at the epoch(s) of observation. Such modelling would also tell us the relationship between the pulse arrival times actually observed, and those that would have been observed in the absence of the scattering medium. Physical modelling is, however, beyond the scope of this paper.

\subsection{Cyclic spectropolarimetry}
We have seen how cyclic spectroscopy gives access to the intrinsic modulation (pulse) profile of the signal, and that this can reveal new structure (figure 2) which is otherwise masked by the effects of scattering. It is the sharp features of the profile -- those which include a large fraction of high-modulation-frequency Fourier components -- which are most affected by the scattering. All the results shown in this paper are based on a signal combination which approximates Stokes-I (recall that our data have not been polarization calibrated). But many pulsars exhibit highly polarized radio emission, and the polarized pulse profiles may be quite complex (van Straten 2006; Johnston et al 2008). For example, there are pulsars where the profile shows rapid transitions between orthogonal, elliptically-polarized states --- usually referred to as ``orthogonal mode jumps''. Such transitions will be strongly affected by any filtering (temporal smearing) of the signal (Karastergiou 2009).  More generally, it is clear that interstellar scattering can have a profound effect on the apparent polarization properties of pulsars at low frequencies (Li and Han 2003; Kramer and Johnston 2008), and we therefore expect the fidelity of polarization profiles to improve substantially when intrinsic profiles, rather than scattered profiles are used.

Furthermore, it has been emphasised by van Straten (2006) that the most accurate pulse-timing requires accurate polarimetry. These are strong motivations to further develop the methods of this paper to encompass cyclic spectropolarimetry. 

\subsection{Covariance of filter coefficients}
In \S4.3.1 we drew attention to the issue of covariance amongst the parameters describing the filter coefficients (or, equivalently, the impulse-response coefficients). The effects of these covariances are not easy to quantify because (i) the total number of parameters needed to describe the filter is very large ($\sim10^4$ in the present case), and (ii) the covariances depend on the properties of both the filter and the pulse profile -- neither of which is known a priori. What is clear, though, is the qualitative point that the actual uncertainty in the filter coefficients can be much larger than the standard deviation for a single parameter taken in isolation. 

We have argued that there are two aspects of the impulse-response functions, seen in figures 7 and 8, that are probably due to parameter covariances. And one of these -- the power near zero delay, scattered to large Doppler-shifts -- is a very strong feature indeed, being evidently well above the noise floor and potentially masking real features of $h(\tau,\omega)$. In other respects cyclic spectroscopy seems to be a near-ideal tool for studying the propagation of radio-pulsar signals, and the issue of parameter covariance consequently deserves further study.

We can identify two aspects that merit particular attention. The first is a thorough understanding of the origin of parameter covariance, and thus how it manifests itself in different representations of the data. Our preliminary analysis (\S4.3.1) suggests that strong covariance can be traced to pure-phase modifications of the filter. That analysis was only carried through for the simplest possible filter model ($H(\nu)=1$), and needs to be revisited using more general models. In cases where the data cannot constrain pure-phase modifications of the filter to be small compared to 1~radian, the problem is akin to one of phase-retrieval. Such problems are notoriously difficult, and the difficulty is associated with non-convexity of the target set (Bauschke, Combettes and Luke 2002).

With an understanding of the origin of the covariances one would be in a good position to tackle the key question of how to mitigate their effects on the filter models. For example, in \S4.3.1 we noted that the well-determined/poorly-determined parameter combinations are sum/difference terms of $h(\tau_j)$ and $h^*(-\tau_j)$, so one might think of enforcing causality in the solutions, such that $h(\tau_j)=0$ for all $\tau_j<0$.

\section{Summary and conclusions}
Cyclic spectroscopy of PSR B1937+21 was undertaken with a 15~second cadence over a 4~MHz  band at 428~MHz, starting from voltages recorded with the Arecibo radio telescope. By least-squares fitting we determined the impulse response function of the ISM for each cyclic spectrum separately, and the intrinsic pulse-profile averaged over the whole observation. In this way we obtained the 428~MHz pulse-profile of B1937+21 free of the influence of interstellar scattering, revealing some weak, but sharp features that had not previously been seen at low radio-frequencies.

From our temporal sequence of impulse response functions we derive the delay-Doppler field image. This image exhibits a noise floor at $-51$~dB relative to the peak power, and we are thus able to see faint features in the angular structure of the received field. Several power concentrations are visible in the delay range $200-400\;{\rm\mu s}$. These concentrations can plausibly be attributed to a single plasma-lens, a few AU in diameter, but alternative interpretations are possible. Regardless of their physical origin, the scattered power concentrations are expected to have a deleterious effect on the pulse-timing experiments that are utilizing this pulsar. To accurately describe and remove these effects it is necessary to have a physical model of the various propagation paths by which the signal reaches the telescope. We did not attempt any physical modelling, but we have shown that cyclic spectroscopy provides us with a large quantity of information on these paths, and thus faciltates that process.

We caution that our fitting procedure is adversely affected by covariance amongst some combinations of the $\sim10^4$ fit parameters. These covariances were identified as the origin of the scattered power artifact in our delay-Doppler image. Parameter covariance appears to be the main challenge currently facing widespread application of cyclic spectroscopy.

\acknowledgments
We thank Dan Stinebring for helpful discussions that prompted our examination of parameter covariances. This paper is dedicated to the memory of Don Backer.



\appendix

\section{Tests of the filter optimization code}
Here we describe tests which we have undertaken to evaluate the performance of our software. Three different aspects of the optimization were compared:  L-BFGS  versus other algorithms; lag-space versus frequency-space optimization; and Unit versus Proximate initializations. All of these comparisons were made using cyclic spectrum samples \#2-11 of PSR B1937+21 recorded at Arecibo on MJD53873 (first processing of the data: see \S6).

We do not expect that our conclusions regarding the relative merits of the different optimization paths are machine dependent. But for reference: the machine used for these tests was a MacBook Pro with a dual core 2.7~GHz Intel processor and 8~GB RAM installed. With this machine almost all algorithms required approximately 2 seconds to complete a single iteration, so run-times for the various approaches can be compared directly from the number of steps required to complete the optimization.

Table 1 sets out the results of our tests. The first three columns show the {\it NLopt\/} algorithm used, the space in which the filter was optimized, and the initialization conditions. Column four shows the average number of steps (rounded to the nearest integer) required to find the best-fit model for the ten sample cyclic spectra.  Column five shows the number of sample cyclic spectra in which a particular configuration yielded the best result (i.e. lowest value of $M_{min}$) out of all of the configurations tested. And the final column shows the average value of $M_{min}$, relative to the best-performing configuration, in units of $\sigma^2$ (rounded to the nearest integer). The ordering of the outcomes in the table was dictated by the results given in the last column, because a high-quality fit is our main objective. In the following sections we consider the outcomes presented in table 1, and their implications for the choice of optimization approach.

\begin{deluxetable}{lccccc}
\tabletypesize{\scriptsize}
\tablecaption{Results of modelling ten sample cyclic spectra.\label{table1}}
\tablewidth{0pt}
\tablehead{
\colhead{{\it NLopt\/}}  &\colhead{Lag or}  & \colhead{Unit or}& \colhead{Avg.}      &\colhead{Best}  & \colhead{$\delta M_{min}$}     \\
\colhead{Algorithm}    & \colhead{Freq.}      & \colhead{Prox.} & \colhead{Step}         &\colhead{\#}     & \colhead{$(\sigma^2)$} }
\startdata
L-BFGS &  Lag     & Prox & 227    & 3  &  0 \\
L-BFGS &  Freq.     & Prox & 262  & 3 &  3 \\
L-BFGS &  Lag     & Unit & 514     & 3 & 8 \\
VarMetric2 & Lag & Prox &   237  &    &  9  \\
L-BFGS &  Freq     & Unit & 680    &    & 10 \\
VarMetric2 & Freq & Prox & 223   &    & 16  \\
VarMetric1 & Lag & Prox &  191   &    & 17   \\
VarMetric1 & Freq & Prox &   221 & 1 & 17   \\
VarMetric2 & Lag & Unit &  499    &     &  29 \\
VarMetric2 & Freq & Unit &   500  &     &  30  \\
VarMetric1 & Freq & Unit &  497   &     &  33  \\
VarMetric1 & Lag & Unit &   471   &     &  39  \\
MMA & Lag             & Prox & 1082&     &  309  \\
TNewtonPR & Lag & Unit  & 2505  &      &   405   \\
MMA & Lag             & Unit & 3394 &     &  600 \\
MMA & Freq             & Prox &  318&     &  616  \\
MMA & Freq             & Unit &1009 &     & 1119
\enddata
\tablecomments{This table summarizes the results of the tests described in this Appendix. Each line represents the outcomes from least-squares modelling of $H(\nu)$, or $h(\tau)$, for 10 sample cyclic spectra of B1937+21. In each case there are approximately $3\times10^6$ degrees of freedom and the total number of parameters in the model is roughly 13,000.}
\end{deluxetable}

\subsubsection{L-BFGS vs other algorithms}
By design the {\it NLopt\/} package makes it possible to switch easily between a variety of different optimization algorithms, and thus to select the best one for the task at hand: to change algorithms is simply a matter of altering one line of code. The algorithms available within {\it NLopt\/} include both global and local methods. Global methods are not practical for our problem because of the large-scale nature of the optimization: it would be necessary to thoroughly search a space of $\sim10^4$ dimensions in order to find the global minimum.

Of the local methods, there are algorithms which require derivatives of $M$ to be supplied, and those which do not. As we are able to supply derivatives, and this is a major advantage in exploring the hypersurface of $M$, we restrict ourselves to those algorithms which make use of the gradient of $M$; there are five such algorithms available in {\it NLopt.\/} One of these, SLSQP (``Sequential Least Squares Quadratic Programming''; Kraft 1994), had not completed a single step after more than an hour of run-time, at which point we terminated the optimization by force. The failure of SLSQP on our optimization problem is not surprising: it uses dense-matrix methods which, for our problem, requires $\sim10^4$ times more storage space and run-time than a limited-memory algorithm. 

Results for the remaining four algorithms are given in table 1. We can see a clear division between these four: the Method of Moving Asymptotes (MMA; Svanberg 2002) and the Truncated Newton method (TNewtonPR; Dembo and Steihaug 1982) both performed poorly on our optimization task, in terms of the quality of fit and run-time, when compared to the Variable Metric (in either rank 1 or rank 2 forms: VarMetric1,2; Vl\v{c}ek and Luk\v{s}an 2006) and L-BFGS algorithms.  We note the failure of TNewtonPR to complete the optimization task from Proximate initialization, or from Unit initialization in frequency-space, hence the omission of those results. It is clear that MMA and TNewtonPR are uncompetitive for our optimization task and we do not consider them further.

It is not surprising that the VarMetric and L-BFGS algorithms yield similar results as they are similar algorithms. Nevertheless, our tests do show a clear preference for L-BFGS over either of the variable metric methods, with L-BFGS providing the three top-performing configurations, as gauged by $\delta M_{min}$, and 9/10 of the best individual fits (column 5 of table 1).

\subsubsection{Lag-space vs. frequency-space}
We have already noted (\S4.2) that lag-space optimization is expected to be superior to a frequency-space approach, because of the traps present in the latter space. This expectation is borne out in practice, with lag-space optimization yielding better fits than the corresponding frequency-space optimization in almost every case in table 1. However, the difference is not very great. We interpret this as meaning that L-BFGS and the VarMetric algorithms obtain enough information on the hyper-surface of $M$ to allow them to avoid most of the traps.

One potential problem which we noticed during our tests is that L-BFGS, when used in frequency-space, would sometimes oscillate as it progressed towards the minimum. This phenomenon was most noticeable with Unit initialization; it appears to be responsible for the 30\%\ extra steps required for L-BFGS-Freq-Unit relative to L-BFGS-Lag-Unit. 

We note that the cyclic spectra used for these tests (see \S6) have typical signal-to-noise ratio greater than unity, for low harmonic numbers, on individual channels. It remains to be seen whether frequency-space optimization remains competitive for cyclic spectra which exhibit low signal-to-noise ratio at all harmonic numbers.

\subsubsection{Variation of initialization}
The algorithms tested here are local methods. That is, they locate a minimum of $M$ in the vicinity of the starting point, but this minimum is not guaranteed to be the global minimum of $M$. The local nature of our solutions is something that readers should be aware of. However, reliably finding the true, global minimum of $M$ in a space with $\sim10^4$ dimensions is a difficult problem which does not seem tractable with the computational technologies currently available. Given the difficulty of finding the true minimum of $M$, it behoves us to examine the sensitivity of our results to the starting point from which the optimization of $H$ proceeds. 

Unsurprisingly, table 1 shows that optimization from a Proximate initialization is roughly a factor of two quicker than from Unit initialization. And Proximate initialization always yields a significantly better fit, for a given choice of algorithm and optimization-space.  Bearing in mind the large-scale nature of the optimization, with $\sim10^4$ parameters, some sensitivity to the initialization conditions is not surprising. 

The fact that there are significant differences between Unit and Proximate initializations suggests the specific question ``how far are our best results from the corresponding global minima?"  As a partial answer to that question we can compare the results of different Proximate initializations, because each of the 10 sample cyclic spectra used in our tests has cyclic spectra taken immediately before and immediately after, and we can step through this sequence in either direction. Referring to the L-BFGS-Lag-Proximate results in table 1 as ``Forward'' initialization, we find that the corresponding ``Backward'' initialization typically gives worse results, with the average $M_{min}$ being larger by $7\sigma^2$ and needing 31 more steps per cyclic spectrum, on average, to complete. Forward initialization produced a better fit than Backward for eight of the ten spectra,\footnote{This level of asymmetry between Forward and Backward initialization is slightly surprising, being expected only once in 18 trials, but we have no explanation other than as a random occurence.} and the root-mean-square difference between the corresponding $M_{min}$ values is approximately $21\sigma^2$. Clearly the variations of the L-BFGS-Lag-Proximate outcomes, relative to the true minimum for each spectrum, must therefore be at least as large as $21\sigma^2$, indicating that there is room for some significant improvement.

This point was confirmed by the following: we ran the whole suite of optimization tests again, but with a tighter fractional tolerance on $M$ of ${\rm0.01/N_{dof}}$ for the stopping criterion. For each of the ten sample cyclic spectra, we took the lowest value of $M_{min}$ (regardless of the configuration which achieved that result) as a reference point. Compared to that reference point, we find that the best-performing configuration of the standard-precision tests (i.e. L-BFGS-Lag-Prox; table 1) is worse by $\delta M_{min}\simeq41\sigma^2$, on average, for each cyclic spectrum.

In the high-precision suite of tests we observed that none of the consistent outcomes of table 1 -- i.e. L-BFGS better than other algorithms, Prox better than Unit, Lag better than Freq -- were reproduced. Not surprisingly, the differences in $M_{min}$ amongst the 12 tested configurations were considerably smaller than shown in table 1, with the worst-performing configuration being only $7\sigma^2$ above the best (cf. $39\sigma^2$ in table 1). These facts suggest that in the high-precision tests all configurations have penetrated well into the noise-limited region of the optimization. The penalty for doing so, of course, is that many more steps are required to achieve that outcome --- 784 steps, on average, for L-BFGS-Lag-Prox, which is more than 3 times the number of steps required to satisfy our usual stopping criterion (see table 1). 

\appendix

\section{Estimation of model uncertainties}

We have already determined the curvature of $M$ with respect to the coefficients describing $S_x$ and $H$ (equations 28 and 29). For the parameters describing the lag-space representation of the filter, the curvatures can be obtained by taking the real and imaginary parts of the relations
\begin{equation}
{{\partial^2 M}\over{\partial h_{rm}\,\partial h_{rj}}}+i\,{{\partial^2 M}\over{\partial h_{rm}\,\partial h_{ij}}}=A_{mj}+C_{mj},
\end{equation}
and
\begin{equation}
{{\partial^2 M}\over{\partial h_{im}\,\partial h_{ij}}}-i\,{{\partial^2 M}\over{\partial h_{im}\,\partial h_{rj}}}=A_{mj}-C_{mj},
\end{equation}
where the matrices $A$ and $C$ are given by
\begin{equation}
A_{mj}={4\over{\rm N_\nu^3}}\!\sum\limits_{n,\alpha\ne0} \!|S_x(\alpha)|^2 \cos\left[2\pi \alpha( \tau_j-\tau_m)\right]  h_n^*\,h_{n+j-m},
\end{equation}
and
\begin{equation}
C_{mj}={4\over{\rm N_\nu^3}}\!\sum\limits_{n,\alpha\ne0} \!\!\;|S_x(\alpha)|^2 \cos\left[2\pi \alpha( \tau_j-\tau_n)\right] h_n\,h_{m+j-n}.
\end{equation}
Here we have used notation such that $h_{n+j-m}$ means $h(\tau_n+\tau_j-\tau_m)$, for example; and we have neglected the contribution from a sum over the residuals, whose expectation is zero.

\subsection{Noise levels for $H(\nu)=1$}
It is clear that the uncertainties in our parameter estimates depend on the filter coefficients and intrinsic pulse profile. But for our purposes here it suffices to determine rough estimates of the parameter uncertainties. To proceed we therefore consider the particular case $H(\nu)=1$. For this circumstance we obtain
\begin{equation}
{{\partial^2 M}\over{\partial S_{rm}^2}} ={{\partial^2 M}\over{\partial S_{im}^2}} =4\, {\rm N_\nu},
\end{equation}
and
\begin{equation}
{{\partial^2 M}\over{\partial H_{rk}^2}}={{\partial^2 M}\over{\partial H_{ik}^2}} =4F^2,
\end{equation}
where $F$ is a measure of the total pulsed flux, with
\begin{equation}
F^2:=\sum\limits_{\alpha\ne0} \;|S_x(\alpha)|^2.
\end{equation}
For the lag representation of the filter we find
\begin{equation}
{{\partial^2 M}\over{\partial h_{rj}^2}}={{\partial^2 M}\over{\partial h_{ij}^2}}={4\over{\rm N_\nu}} F^2, \quad (\tau_j\ne0),
\end{equation}
and for $\tau_j=0$ the curvature with respect to the real part of the coefficient $h_j$ is twice this value, whereas there is no curvature with respect to the imaginary part. This last point, which implies a formally infinite uncertainty, should not cause concern because the overall phase of the filter is completely arbitrary.

Using equation 39 we can immediately translate these curvatures into standard deviations. The results are
\begin{equation}
\delta S_m = {\sigma\over\sqrt{2\rm N_\nu}},
\end{equation}
\begin{equation}
\delta H_k = {\sigma\over{F\sqrt{2}}},
\end{equation}
and
\begin{equation}
\delta h_j = {\sigma\over F}\sqrt{{\rm N_\nu}\over{2}}, \quad (\tau_j\ne0).
\end{equation}
In all these cases the coefficients are complex; the quoted uncertainty is the uncertainty in the real part of the coefficient, which is equal to the uncertainty in the imaginary part. With the exception of one coefficient of $h$, the standard deviation is uniform across each set of coefficients.

In practice the system noise, $\sigma$, is dependent on the total number of radio-frequency channels, ${\rm N_\nu}$, because we have a fixed total bandwidth, $B$, for the instrument. Thus ${\rm N_\nu} \Delta\nu =B$, and equation 14 can be written
\begin{equation}
\sigma ={\cal S}_{sys}\sqrt{ {\rm N_\nu} \over {B\,\Delta t } }.
\end{equation}
A further simplification is appropriate. For cyclic spectroscopy of a pulsar with period $P$, the pulsar's rotation frequency $\Omega=1/P$ is necessarily equal to the spacing in modulation frequency, $\Delta \alpha$, and in turn this is the natural choice for channelisation, $\Delta\nu$. Thus the natural configuration is $PB={\rm N_\nu}$, and for this circumstance we obtain
\begin{equation}
\delta S_m = {{\cal S}_{sys}\over\sqrt{2B\, \Delta t} },
\end{equation}
\begin{equation}
\delta H_k ={ {\cal S}_{sys}\over {F}}\sqrt{P \over{2 \Delta t}  },
\end{equation}
and
\begin{equation}
\delta h_j ={ {{\cal S}_{sys}P}\over {F}}\sqrt{B \over{2 \Delta t}  }, \quad (\tau_j\ne0).
\end{equation}

\subsection{Noise levels for more general filters}
The curvature of the demerit function with-respect-to the various model parameters depends on the structure in the filter functions, as manifest in equations 28, 29, A3, A4, but we have so far considered only the simplest filter, $H(\nu)=1$. We now consider how structure in the filter affects the noise level on various parameters.

It is, of course, possible to concoct bizarre examples of filters which imply correspondingly unusual noise properties. But we shall ignore such possibilities as our purpose here is to describe what one might normally expect to encounter in practice. To that end we will restrict our discussion to cases where $\langle|H(\nu)|^2\rangle\sim\langle|H(\nu)|^4\rangle\sim1$, and we will characterize the impulse response function by a typical scattering time, $\tau_s$, corresponding to a filter decorrelation bandwidth $\sim1/\tau_s$.

Consider first the noise level for the pulse harmonic coefficients. For low harmonics the summation in equation 28 is approximately ${\rm N}_\nu\langle|H(\nu)|^4\rangle$. But at higher harmonics, where $|\alpha_m|\tau_s\sim1$, there is some decorrelation between $|H(\nu-\alpha_m/2)|$ and $|H(\nu+\alpha_m/2)|$ and the sum declines. In the limit of complete decorrelation, $|\alpha_m|\tau_s\gg1$, the summation yields ${\rm N}_\nu\langle|H(\nu)|^2\rangle^2$. Providing that both second- and fourth-order expectation values are of order unity, this is not a big effect. For example, in the random-phasor picture for the electric field the intensity statistics are exponential, so $\langle|H(\nu)|^2\rangle=1$ and $\langle|H(\nu)|^4\rangle=2$, yielding a noise level for high harmonics which is $\sqrt{2}$ larger than for low harmonics. In this picture, the noise level for high harmonics coincides with the value quoted in equation A13, for the case $H(\nu)=1$.

Quite a different situation arises for the filter coefficients $H_k$. It is evident that the curvatures given in equation 29 may be much less than $4F^2$ in regions where the filter function is small, with correspondingly large errors on those coefficients. As with the noise on the pulse harmonics,  there are two different limiting cases relating to the value of the typical scattering time. Most of the pulsed flux, $F$, is contributed by harmonics up to $|\alpha_m|\sim1/w$, where $w$ is the temporal width of the pulse. If $\tau_s\ll w$ then the filter function $H(\nu_k-\alpha)$ is almost constant over the range of $\alpha$ which contributes most to $F$, so the curvature in equation 29 becomes $4F^2|H_k|^2$. Clearly this curvature could be very large (small) in comparison with the estimate given in equation A6, leading to correspondingly small (large) errors in the $H_k$ estimates. In the opposite limit, where $\tau_s\gg w$, the filter coefficient $|H(\nu-\alpha)|$ changes rapidly with harmonic number and we obtain a curvature  estimate $\sim 4F^2\langle|H(\nu)|^2\rangle\sim4F^2$, comparable to that given in equation A6. 

Finally we consider the effect of a structured filter on the errors associated with the lag-space filter coefficients, $h_j$. The curvatures of the merit function with respect to real and imaginary parts are (equations A3 and A4) made up of two terms. The first term is the same in both cases and we expect it to be $4F^2 \langle|H(\nu)|^2\rangle/{\rm N}_\nu\sim4F^2/{\rm N}_\nu$. The second term differs in sign between the real and imaginary parts of the coefficients; it is the real part of a sum of complex numbers. In normal circumstances those complex numbers bear no particular phase relationship to each other, so the second term is typically small in comparison with the first. We therefore neglect it, and we conclude that in normal circumstances the curvatures given in equation A8 are appropriate to all lag-space filter coefficients.

\end{document}